 \documentclass[journal]{IEEEtran}

\usepackage{cite}
\usepackage{amsmath,amssymb,amsfonts}
\usepackage{algorithm}  
\usepackage{algpseudocode}  
\usepackage{graphicx}
\usepackage{epstopdf}
\usepackage{textcomp}
\usepackage{xcolor}
\usepackage{bm}
\usepackage{enumerate}
\usepackage{amsmath}
\ifCLASSINFOpdf
\else
\fi
\begin{document}
	
%
\title{Analysis and Algorithm for Multi-IRS Collaborative Localization via Hybrid Time–Angle Estimation}
%
%
%

\author{Ziheng~Zhang,~Wen~Chen,~Qingqing~Wu,~Haoran~Qin,~Zhendong~Li, and Qiong~Wu


\thanks{Z. Zhang, W. Chen, Q. Wu and H. Qin are with the Department of Electronic Engineering, Shanghai Jiao Tong University, Shanghai 200240, China (e-mail: zhangziheng@sjtu.edu.cn; wenchen@sjtu.edu.cn; 
qingqingwu@sjtu.edu.cn;haoranqin@sjtu.edu.cn)}
\thanks{Z. Li is with the School of Information and Communication Engineering, Xi'an Jiaotong University, Xi'an 710049, China (email: lizhendong@xjtu.edu.cn).}
\thanks{Q. Wu is with the School of Internet of Things Engineering, Jiangnan
University, Wuxi, China (emali: qiongwu@jiangnan.edu.cn).}


\thanks{(\emph{Corresponding author: Wen Chen.})}}
\maketitle
\begin{abstract}
This paper proposes a novel multiple intelligent reflecting surfaces (IRSs) collaborative hybrid localization system, which involves deploying multiple IRSs near the target area and achieving target localization through joint time delay and angle estimation. Specifically, echo signals from all reflective elements are received by each sensor and jointly processed to estimate the time delay and angle parameters. Based on the above model, we derive the Fisher Information Matrix (FIM) for cascaded delay, Angle of Arrival (AOA), and Angle of Departure (AOD) estimation in  semi-passive passive models, along with the corresponding Cramér-Rao Bound (CRB). To achieve precise estimation close to the CRB, we design efficient algorithms for angle and location estimation. For angle estimation, reflective signals are categorized into three cases based on their rank, with different signal preprocessing. By constructing an atomic norm set and minimizing the atomic norm, the joint angle estimation problem is transformed into a convex optimization problem, and low-complexity estimation of multiple AOA and AOD pairs is achieved using the Alternating Direction Method of Multipliers (ADMM). For location estimation, we propose a three-stage localization algorithm that combines weighted least squares, total least squares, and quadratic correction to handle errors in the coefficient matrix and observation vector, thus improving accuracy. Numerical simulations validate the superiority of the proposed system, demonstrating that the system's collaboration, hybrid localization, and distributed deployment provide substantial benefits, as well as the accuracy of the proposed estimation algorithms, particularly in low signal-to-noise ratio (SNR) condition.

\end{abstract}


%
\IEEEpeerreviewmaketitle

\section{Introduction}
The integration of sensing technologies into 6G networks is expected to be a defining characteristic, enhancing the capabilities of wireless communication systems in unprecedented ways \cite{10012421}. This integration primarily involves two key aspects: one is the emergence of environment-aware applications, such as autonomous vehicles \cite{10330577}, drone delivery systems \cite{10098686}, and advanced satellite communications \cite{10768935}, and the other is dual role of communication signals for both data transmission and sensing, allowing for target localization and environmental imaging \cite{9945983}. Additionally, environmental data gathered through sensing can improve channel estimation and reduce the need for pilot signals, enhancing overall network efficiency \cite{10630588}. However, delivering high-precision sensing capabilities across diverse and complex electromagnetic environments presents several challenges. Firstly, high-frequency sensing signals, required for precise localization, suffer from significant propagation loss due to free-space path loss and atmospheric attenuation, limiting their effective range \cite{10416896}. Secondly, achieving reliable line-of-sight (LoS) links between base stations (BS) and targets is particularly difficult in urban environments, where obstacles obstruct signal paths. Finally, the computational demands for processing large-scale real-time data, such as radar imaging or MIMO measurements, pose significant challenges in terms of both computational power and energy consumption. Overcoming these challenges is essential for improving the performance of sensing applications in 6G networks.

Intelligent Reflecting Surface (IRS) has emerged as a transformative technology in the development of 6G wireless networks, offering the ability to manipulate the electromagnetic environment in a cost-effective and energy-efficient manner. By dynamically adjusting the phase shifts of reflected signals, IRS enhances communication performance, mitigates signal blockage, and improves spectral efficiency. This capability extends beyond conventional communication systems and has led to growing interest in IRS-assisted wireless sensing, where it facilitates the creation of additional LoS links and enhances environmental sensing \cite{10555049}. In the context of 6G, which is characterized by its focus on ubiquitous intelligence and the integration of sensing and communication (ISAC), IRS plays a crucial role in advancing sensing capabilities. Specifically, IRS can significantly improve target detection, localization, and environmental mapping by optimizing signal reflection paths, particularly in non-line-of-sight (NLoS) environments \cite{9724202}. Recent research has proposed several IRS-assisted sensing architectures, including fully passive IRS sensing \cite{10547700}, semi-passive IRS sensing \cite{10547700}, and IRS-based self-sensing systems \cite{9340586}. These architectures differ in terms of the control and feedback mechanisms employed.

Under these IRS-assisted sensing architectures—including fully passive, semi-passive, and IRS-based self-sensing systems—that effectively exploit IRS's ability to dynamically reconfigure electromagnetic environments and establish additional LoS links \cite{10585319,10289119,10279612}, recent research has developed innovative approaches for wireless localization, target tracking, and detection applications \cite{9777939,10703129,10133053,9963716,10443321,10634199,10575930,zhang2024full,10464564,9725255}. In wireless localization, extensive studies have shown that IRS can overcome inherent limitations of conventional locationing methods in complex environments \cite{10844684,9777939,10703129}. Specifically, IRS establishes virtual line-of-sight (VLoS) links by manipulating reflected signal paths, thus compensating for direct-path blockage and multipath interference \cite{9777939,10464353}. This capability not only mitigates challenges posed by obstacles but also strengthens signal availability, particularly beneficial in massive MIMO systems \cite{10703129, 10133053}. Consequently, IRS-aided localization exhibits improved accuracy and stability compared to traditional schemes, \cite{9777939,10703129,10133053}.
Similarly, IRS demonstrates substantial potential in target tracking, addressing typical accuracy deterioration arising from multipath fading and signal attenuation in dynamic multi-target scenarios \cite{9963716,10443321,10634199,10575930}. By adaptively optimizing reflection phases and beam patterns, IRS provides stable and reliable tracking paths, reducing tracking error and ensuring continuous signal availability \cite{9963716,10443321,10575930}. Such adaptability enables effective real-time optimization of propagation paths, significantly improving multi-target tracking accuracy under severe NLoS conditions \cite{10634199,10575930}.
Furthermore, IRS has enabled new capabilities in target detection, especially where traditional systems suffer from limited coverage and insufficient sensitivity \cite{zhang2024full,10464564,9725255}. By adaptively controlling reflected signals, IRS enhances echo signals from targets, thereby improving detection sensitivity in cluttered and obstructed environments \cite{10464564}. This functionality is particularly advantageous for real-time monitoring in scenarios such as intelligent transportation and autonomous driving, as IRS effectively compensates for direct-path losses and substantially boosts system detection capabilities \cite{zhang2024full,10464564,9725255}.
Looking forward, further research is anticipated to focus on collaborative multi-IRS strategies, intelligent beam management schemes, and comprehensive optimization in more complex scenarios, thereby achieving superior sensing performance and precision in future wireless sensing systems.

{\color{blue}
First, while IRS technology has shown great promise, the study of multi-IRS collaboration remains insufficient. Although prior works have explored the benefits of multi-IRS systems in extending coverage, they primarily focus on non-cooperative schemes where each IRS serves a separate zone \cite{10497119}, or employ time-division protocols to avoid inter-IRS interference \cite{10506632,11085137}. A framework enabling fully collaborative localization at the physical layer, where multiple semi-passive IRSs jointly process all signals, including inter-IRS reflection paths, remains to be thoroughly investigated. Secondly, there is a noticeable gap regarding the joint use of sensing information. Existing IRS-assisted localization methods predominantly rely on a single type of sensing information. For instance, a significant body of research focuses on Angle-of-Arrival (AOA)-based localization \cite{10497119,10506632,11085137,10506936}, employing advanced signal processing techniques like MUSIC, ESPRIT \cite{10506632}, or Atomic Norm Minimization (ANM) \cite{10621068,10443910,10540064} to achieve high-resolution AOA estimation. However, these angle-only methods neglect the valuable range information embedded in time-delay measurements. While some works have separately studied Time-of-Arrival (TOA)-based localization \cite{10747182}, the potential of a hybrid approach that jointly processes both angle and delay information within a multi-IRS collaborative framework to maximize localization accuracy has not yet been fully explored.
Finally, the research on specific signal processing techniques for estimating these parameters remains limited. Most existing research on IRS-assisted sensing focuses primarily on system-level optimization, such as joint beamforming design to minimize the Cramér-Rao Bound (CRB) \cite{10497119,11085137}. There is insufficient investigation into the signal processing algorithms for extracting essential information (such as angle and delay) from the reflected signals as a prerequisite for localization. Although efficient algorithms leveraging angular domain sparsity have been proposed for channel estimation \cite{9760391}, the design of a complete set of parameter estimation algorithms specifically for the hybrid localization scenario would not only enhance accuracy but also provide valuable insights for system and protocol design, ultimately contributing to the overall performance improvement of IRS-assisted sensing systems. }

Motivated by the above considerations, we study a multiple semi-passive IRS deployed near the target area to enable collaborative localization of potential targets, as shown in Fig. 1. Our objective is to analyze the CRB of the proposed system for target localization and design efficient algorithms to estimate the cascaded delay, angle, and the target's location. It is worth noting that collaborative localization refers to the unified signal processing of the reflections received by the sensors of all semi-passive IRS, while hybrid localization involves jointly estimating the location based on both delay and angle information. {\color{blue}The analysis and algorithms in this paper are specifically developed for a single-target localization scenario, which lays a crucial foundation for more complex multi-target systems.} The main contributions of the paper are summarized as follows.
{\color{blue}
\begin{itemize}
\item  \textbf{Proposed a multi-IRS collaborative localization system model  and derived its performance bounds.} We propose a novel multi-IRS collaborative hybrid localization system. A key assumption of this model is the mutual orthogonality of signals reflected by different IRSs, which establishes a tractable theoretical framework for designing our core estimation algorithms. The feasibility of this idealized assumption is grounded in established techniques, such as zero-forcing (ZF) beamforming at the base station or the strategic orthogonal deployment of IRSs. Based on this model, we derive the Fisher Information Matrix (FIM) for cascaded delay, AOA, and Angle of Departure (AOD), and further obtain the CRB for target localization.

\item \textbf{Designed a joint angle estimation algorithm adaptable to various signal conditions.}
For the angle estimation problem, we classify the IRS reflected signal into three distinct cases based on its matrix rank and design unique signal preprocessing techniques for each. By constructing an atomic norm set, we transform the joint angle estimation problem into a convex optimization problem. We then derive angle estimation algorithms for different cases using the Alternating Direction Method of Multipliers (ADMM), achieving low-complexity estimation of multiple AOA and AOD pairs.

\item \textbf{Designed a robust three-stage hybrid localization algorithm.} For the location estimation problem, we propose a three-stage algorithm to perform hybrid localization that is resilient to errors in delay and angle estimates. By combining Weighted Least Squares (WLS), Total Least Squares (TLS) to handle errors in the coefficient matrix, and a final correction step using second-order information, the algorithm systematically addresses error propagation and accumulation, yielding an accurate location estimate .

\item \textbf{Validated system gains and algorithm performance under the ideal model via simulations.}
Finally, the superiority of the proposed system is validated through numerical simulations. The results clearly demonstrate the significant performance gains from multi-IRS collaboration, the use of hybrid information, and the distributed deployment strategy under the ideal condition of signal orthogonality. The simulations also confirm the effectiveness and accuracy of the proposed joint angle estimation and three-stage localization algorithms, particularly in low signal-to-noise ratio (SNR) environments.
\end{itemize}}

The rest of this paper is as follows. In Section II, we introduce the system model of multiple IRSs collaborative hybrid localization. Then, Section III  the CRB for delay, angle, and location estimation for semi-passive systems. Then, in Sections IV and V,  we elaborate on the algorithms for angle and location estimation, respectively. Section VI reveals the performance superiority of the proposed system and algorithm compared to other benchmarks and some useful insights. Finally, Section VII concludes this paper.

\textit{Notations:} Matrices, vectors and scalars
are represented by bold uppercase, bold lowercase and standard lowercase letters, respectively. For a complex-valued scalar $x$, $\left| {x} \right|$ denotes its absolute value. For a complex-valued vector $\bf{x}$, ${{\left\| \bf{x}  \right\|}_{p}}$, ${{\left[ \mathbf{x} \right]}_{i}}$ represents the the its $p$-norm and $i$-th element, respectively. For a general matrix $\bf{A}$, $\text{rank}(\bf{A})$, ${\bf{A}}^H$, ${{\mathbf{A}}^{\dagger}}$
 and ${{\left[\bf{A} \right]}_{i,j}}$ denote its rank, conjugate transpose, Moore-Penrose pseudo-inverse and $\left( i,j \right)$-th element, respectively. For a square matrix $\bf{X}$,  and $\text{tr}(\bf{X})$ denote its trace. ${\bf{X}} \succeq 0$ denotes that $\bf{X}$ is a positive semidefinite matrix. ${\mathbb{C}^{M \times N}}$ represents the ${M \times N}$ dimensional complex matrix space. $\mathbb{E}\left( \cdot  \right)$ denotes the expectation operation. $\sim$ represents
“distributed as” and $\mathcal{C}\mathcal{N}\left( {\mathbf{x},\mathbf{R}} \right)$ represents
the distribution of a circularly symmetric complex Gaussian random vector with mean vector $\mathbf{x}$ and covariance matrix $\mathbf{R}$. $\otimes $ represents the Kronecker product.

\section{System Model}
As shown in Fig. 1, the model of a multi-IRSs collaborative hybrid localization system employing a semi-passive architecture is illustrated. This semi-passive architecture for IRS-assisted sensing consists of a BS with $N_t$ transmit antennas for transmitting sensing signals, and $K$ IRSs. Each IRS is equipped with $N$ reflective elements to construct VLoS links. In this semi-passive configuration, the sensors are co-located with the IRSs, meaning each of the $K$ IRSs is equipped with $M$ sensors. {\color{blue}
The location of BS, $k$-th IRS, and target are denoted by ${{\mathbf{l}}_{\text{BS}}}=\left( {{x}_{\text{BS}}},{{y}_{\text{BS}}}\right)$, ${{\mathbf{l}}_{\text{I},k}}=\left( {{x}_{\text{I},k}},{{y}_{\text{I},k}}\right)$ and ${{\mathbf{x}}}=\left( {{x}_{\text{T}}},{{y}_{\text{T}}}\right)$, respectively. The distance and angle between the target and the $k$-th IRS can be expressed as ${{d}_{k}}=c{{\tau }_{k}}=\sqrt{{{\left( {{y}_{\text{T}}}-{{y}_{\text{I},k}} \right)}^{2}}+\left( {{x}_{\text{T}}}-{{x}_{\text{I},k}} \right)^{2}}$ and ${{\theta }_{k}}=\arctan \left( {{{y}_{\text{T}}}-{{y}_{\text{I},k}}}/{{{x}_{\text{T}}}-{{x}_{\text{I},k}}} \right)$, respectively where $c$ and ${{\tau }_{k}}$ represent the speed of light and corresponding propagation delay, respectively.} Under the assumption of the far-field condition, the AOD from the reflective elements on IRS to target is the same as the AOA, i.e.,  ${{\theta }_{k}}={{\theta }_{k,\text{AOA}}}={{\theta }_{k,\text{AOD}}}$.

\begin{figure}
\centerline{\includegraphics[width=9cm]{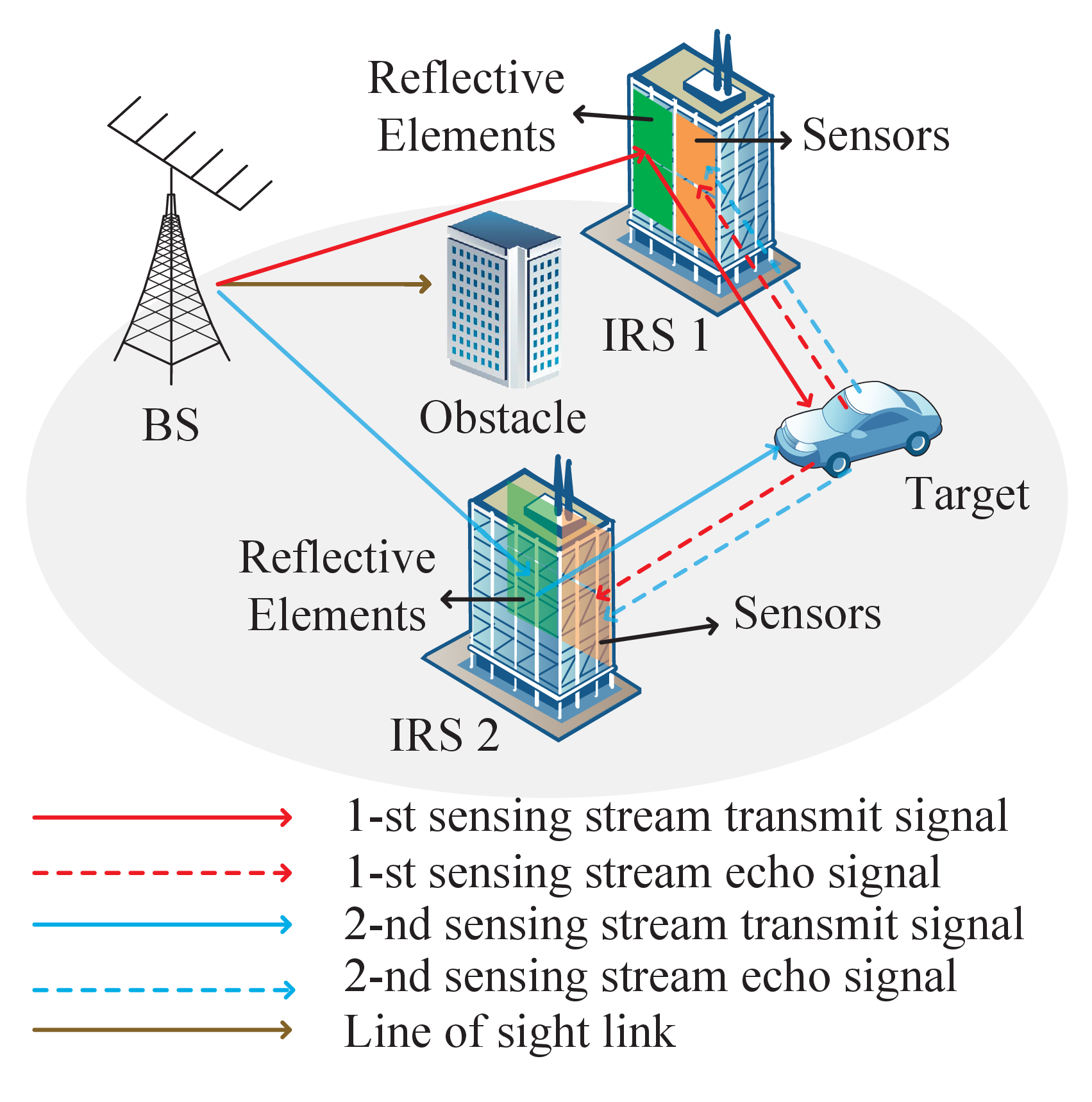}}
	\caption{Multiple IRSs collaborative hybrid localization system.}
	\label{Fig1}
\end{figure}

\subsection{Channel Model}
The channel between the BS and the $k$-th IRS ${{\mathbf{H}}_{k}}$ consists of $R_k$ multipath components, which can be expressed as
{\color{blue}
\begin{align}
{{\mathbf{H}}_{k}}=\sum\limits_{r=1}^{{{R}_{k}}}{{{\alpha }_{k,r}}\mathbf{a}\left( {{\varphi }_{k,r}},N \right){{\mathbf{a}}^{H}}\left( {{\theta }_{k,r}},{{N}_{t}} \right)},
\end{align}}
where ${r}_{k}$ represents the number of paths between the base station and the $k$-th IRS, and ${{\alpha }_{k,r}}$ represent the path gain and 
${{\theta }_{k,r}},{{\varphi }_{k,r}}$ represents the AOD and AOA of the $r$-th path, respectively;  
$\mathbf{a}\left( {\theta_0},{N_0}\right)$ represents the steering vector where ${\theta_0 }$ represents the phase difference between two adjacent elements or antennas, and ${N_0}$ represents the number of antennas transmitted or received, which is defined as
\begin{align}
\mathbf{a}\left( {\theta_0 },{N_0} \right)={{\left[ 1,{{e}^{j2\pi \sin {\theta_0 }}},\cdots {{e}^{j2\pi \left( {N_0}-1 \right)\sin {\theta_0}}} \right]}^{T}}\in {{\mathbb{C}}^{{N_0}}}.
\end{align}

Because localization related information can only be extracted from LoS, the channel from $k$-th IRS to $l$-th IRS via target of semi-passive system can be modeled as 
\begin{align}
\mathbf{H}_{l,k}^{}={{\alpha }_{l,k}}\mathbf{a}\left( {{\theta }_{l,\text{AOA}}},M \right){{\mathbf{a}}^{H}}\left( {{\theta }_{k,\text{AOD}}},N \right),  
\end{align}
where ${{\alpha }_{l,k}}=\sqrt{{{{\lambda }^{2}}\kappa }/{64{{\pi }^{3}}d_{k}^{2}d_{l}^{2}}}$ denotes the path loss, where $\kappa $ denotes the radar cross section, $\lambda $ denotes the wavelength of the signal.

\subsection{Performance Analysis}
Let $\mathbf{X}\in {{\mathbb{C}}^{{{N}_{t}}\times L}}$ be the sensing signal matrix, with $L\gg {{N}_{t}}$ being the length of the radar pulse frame and $\mathbf{X}$ can be modeled as 
\begin{align}
\mathbf{X}=\sum\limits_{k=1}^{K}{{{\mathbf{W}}_{k}}{{\mathbf{X}}_{k}}},
\end{align}
where ${{\mathbf{X}}_{k}},{{\mathbf{W}}_{k}}$ represent the $k$-th sensing stream and its beamforming matrix.   To ensure that the receiver can distinguish the signals corresponding to different paths reflected by the $K$ IRSs, the signals reflected by the $K$ IRSs are designed to be mutually orthogonal,i.e., 
\begin{align}
{{\mathbf{H}}_{k}}{{\mathbf{W}}_{k'}}=0,\forall k\ne k', 
\end{align}
which can be achieved through the orthogonal deployment of IRSs or zero-forcing beamforming. The sensing streams corresponding to the same IRS are assumed to be independent in different pulse frame while the sensing streams corresponding to the different IRSs are assumed to be independent in any frame, i.e.,
\begin{align}\label{signal_orth}
{{\mathbf{X}}_{k}}\mathbf{X}_{k}^{H}={{\mathbf{I}}_{{{N}_{t}}}},{{\mathbf{X}}_{k}}\mathbf{X}_{k'}^{H}={{\mathbf{0}}_{{{N}_{t}}}},\forall k\ne k'.
\end{align}

{\color{blue}We adopt a semi-passive IRS architecture. While this increases hardware complexity, it is a well-established approach in sensing literature that significantly improves the signal-to-noise ratio (SNR) by reducing round-trip path loss . For tractability, our model focuses on the primary reflection paths, assuming sensor data is collected via an ideal backhaul. Higher-order reflections, such as inter-IRS paths, are considered negligible due to higher path loss but can be explored in future work. }For the semi-passive architecture, the signal received by the sensors on the $l$-th IRS can be expressed as
\begin{align}\label{semirl}
{{\mathbf{R}}_{l}}\left( t \right)=\sum\limits_{k=1}^{K}{{{\mathbf{H}}_{l,k}}{{\mathbf{\Theta }}_{k}}{{\mathbf{H}}_{k}}\mathbf{X}\left( t-{{{\dot{\tau }}}_{l,k}} \right)}+{{\mathbf{N}}_{l}},
\end{align}
where ${{\mathbf{N}}_{l}}$ represents the Gaussian white noise at the sensors in $l$-th IRS and $\text{vec}\left( {{\mathbf{N}}_{l}} \right)\sim\mathcal{C}\mathcal{N}\left( 0,\sigma _{l}^{2}{{\mathbf{I}}_{ML}} \right)$, ${{\dot{\tau }}_{l,k}}\triangleq {{\tau }_{\text{B2I,}k}}+{{\tau }_{l,k}}$, ${{\tau }_{\text{B2I,}k}}$ represents the signal propagation delay between the BS and the $k$-th IRS and ${{\tau }_{l,k}}\triangleq {{\tau }_{k}}+{{\tau }_{l}}$ represent the signal cascade propagation delay from $k$-th IRS to the $l$-th IRS via target. {\color{blue} Let ${{\mathbf{\Theta }}_{k}}=\text{diag}\left( {{\alpha }_{k,1}}{{e}^{j{{\beta }_{{k},1}}}},\cdots,{{\alpha }_{k,N}}{{e}^{j{{\beta }_{k,N}}}}\right)$ denotes the reflection coefficient matrix of $k$-th IRS, where ${{\alpha }_{k,n}}$ and ${{\beta }_{k,n}}$ denotes the amplitude and phase reflection coefficient of the $n$-th element of $k$-th IRS, respectively. The phase shift matrix ${{\mathbf{\Theta }}_{k}}$ of each IRS can be dynamically configured to enhance system performance. It is important to note that the performance analysis and estimation algorithms developed in this work are general and applicable to any given IRS phase configuration.}

Since the location of the BS and IRS are fixed, ${\tau }_{\text{B2I,}k}$ can be treated as known. Our hybrid localization model first extracts unknown angle $\boldsymbol\theta =\left[ {{\theta }_{1}},{{\theta }_{2}},\cdots, {{\theta }_{K}} \right]\in {{\mathbb{R}}^{K\times 1}}$ and cascade delay $\boldsymbol{\tau }=\left[ {{\tau }_{1,1}},{{\tau }_{1,2}},\cdots ,{{\tau }_{K,K}} \right]\in {{\mathbb{R}}^{{{K}^{2}}\times 1}}$ information from received signal ${{\mathbf{R}}}=\left[ {{\mathbf{R}}_{1}},\cdots {{\mathbf{R}}_K} \right]\in {{\mathbb{C}}^{M\times KL}}$, and then utilizes above hybrid information to estimate the target's location $\mathbf{x}$.

\section{Performance Analysis of Multi-IRS Collaborative Hybrid Localization System}
CRB provides a lower bound of variance of any unbiased estimator, therefore, we first derive the CRB for the location related estimation in the proposed system to characterize the system's performance and validate the effectiveness of the subsequent estimation algorithms.
For any variable to be estimated $\mathbf{u}$, the variance of the unbiased estimator ${{\widehat{\mathbf{u}}}}$ satisfies
\begin{align}
\operatorname{var}\left( \widehat{\mathbf{u}} \right)\ge {{\text{CRB}}}\left( \mathbf{u} \right)={{\mathbf{F}}^{-1}}\left( \mathbf{u} \right),
\end{align}
where ${\text{CRB}}\left( \mathbf{u} \right)$ and ${{\mathbf{F}}^{-1}}\left( \mathbf{u} \right)$ denote the CRB matrix and FIM of $\mathbf{u}$ to be estimated, respectively. 
\subsection{CRB for Cascade Delay Estimation}\label{crb_angle}
In this subsection, we analyze the CRB for $\boldsymbol{\tau }$ estimation involved in the localization system. The Fisher matrix for 
$\boldsymbol{\tau }$ can be modeled as
\begin{align}
\resizebox{0.88\hsize}{!}{$\mathbf{F}\left( \mathbf{\tau } \right)=\left[ \begin{matrix}
   F\left( {{\tau }_{1,1}},{{\tau }_{1,1}} \right) & F\left( {{\tau }_{1,1}},{{\tau }_{1,2}} \right) & \cdots  & F\left( {{\tau }_{1,1}},{{\tau }_{K,K}} \right)  \\
   F\left( {{\tau }_{1,2}},{{\tau }_{1,1}} \right) & F\left( {{\tau }_{1,2}},{{\tau }_{1,2}} \right) & \cdots  & F\left( {{\tau }_{1,2}},{{\tau }_{K,K}} \right)  \\
   \cdots  & \cdots  & \cdots  & \cdots   \\
   F\left( {{\tau }_{K,K}},{{\tau }_{1,1}} \right) & F\left( {{\tau }_{K,K}},{{\tau }_{1,2}} \right) & \cdots  & F\left( {{\tau }_{K,K}},{{\tau }_{K,K}} \right)  \\
\end{matrix} \right],$}
\end{align}
where the element of $\mathbf{F}\left(\boldsymbol{\tau }\right)$ is given by
\begin{align}
F\left( {{\tau }_{l,k}},{{\tau }_{{l}',{k}'}} \right)={{\mathbb{E}}_{\mathbf{R}|\boldsymbol{\tau },\boldsymbol{\theta}}}\left[ \frac{{{\partial }^{2}}\log p\left( \mathbf{R}|\boldsymbol{\tau },\boldsymbol{\theta} \right)}{\partial {{\tau }_{l,k}}\partial {{\tau }_{{l}',{k}'}}} \right],
\end{align}
where $p\left( \mathbf{R}|\boldsymbol{\tau },\boldsymbol{\theta }  \right)$ is the joint probability density function of ${\mathbf{R}}$ 
conditioned on $\boldsymbol{\tau },\boldsymbol{\theta}$ and can be given as 
\begin{equation}\label{pr}
p\left( \mathbf{R}|\boldsymbol{\tau },\boldsymbol{\theta}  \right)=\exp \left( \sum\limits_{l=1}^{K}{\int_{T}{-}\frac{1}{\sigma _{l}^{2}}{{\left| {{\mathbf{R}}_{l}}\left( t \right)-\overline{{{\mathbf{R}}_{l}}}\left( t \right) \right|}^{2}}dt} \right),
\end{equation}
where $\overline{{{\mathbf{R}}_{l}}}\left( t \right)$ denotes the deterministic component of the received echo signal \cite{5466526}. By substituting (\ref{semirl}) into (\ref{pr}) and ignoring terms that are unrelated to the derivative, $F\left( {{\tau }_{l,k}},{{\tau }_{{l}',{k}'}} \right)$ can be expressed as
\begin{equation}
F\left( {{\tau }_{l,k}},{{\tau }_{{l}',{k}'}} \right)=M\alpha _{l,k}^{2}\left\| {{\mathbf{a}}^{H}}\left( {{\theta }_{k,\text{AOD}}},N \right){{\mathbf{\Theta }}_{k}}{{\mathbf{H}}_{k}}{{\mathbf{W}}_{k}}{{{\mathbf{\dot{X}}}}_{k}}\left( t \right) \right\|_{2}^{2}
\end{equation}
when $l=l',k=k'$ and $0$ otherwise, where ${{\mathbf{\dot{X}}}_{k}}\left( t \right)={\partial {{\mathbf{X}}_{k}}\left( t \right)}/{\partial t}$ and the CRB for the cascade delay estimation can be expressed as ${{\text{CRB}}}\left( \boldsymbol{\tau }  \right)={{\mathbf{F}}^{-1}}\left( \boldsymbol{\tau }  \right)$.

\subsection{CRB for Angle Estimation}\label{crb_angle}
In this subsection, we analyze the CRB for angle estimation involved in the localization system, which includes two types of angles: AOA and AOD. 
Similar to the Fisher matrix for delay estimation, the Fisher matrix for AOA and AOD angle estimations can also be represented as a diagonal matrix as 
\begin{align}
& \mathbf{F}\left( {{\boldsymbol{\theta }}_{\text{AOA}}} \right)=\text{diag}\left( F\left( {{\theta }_{1,\text{AOA}}} \right),\cdots ,F\left( {{\theta }_{K,\text{AOA}}} \right) \right), \\ 
& \mathbf{F}\left( {{\boldsymbol{\theta }}_{\text{AOD}}} \right)=\text{diag}\left( F\left( {{\theta }_{1,\text{AOD}}} \right),\cdots ,F\left( {{\theta }_{K,\text{AOD}}} \right) \right),
\end{align}
where $F\left( {{\theta }_{l,\text{AOA}}} \right)$ and  $F\left( {{\theta }_{k,\text{AOD}}} \right)$  can be obtained by computing the second-order derivatives of $p\left( \mathbf{R}|\boldsymbol{\tau },\boldsymbol{\theta}  \right)$ with respect to ${{\theta }_{l,\text{AOA}}}$  and ${{\theta }_{k,\text{AOD}}}$, respectively,i.e., 
\begin{align}
& F\left( {{\theta }_{l,\text{AOA}}} \right)=\sum\nolimits_{k=1}^{K}{\left\| \mathbf{\dot{a}}\left( {{\theta }_{l,\text{AOA}}},N \right)\mathbf{b}_{k}^{H} \right\|_{2}^{2}}, \\ 
 & F\left( {{\theta }_{k,\text{AOD}}} \right)=\sum\nolimits_{l=1}^{K}{\left\| \mathbf{a}\left( {{\theta }_{l,\text{AOA}}},N \right)\mathbf{c}_{k}^{H} \right\|_{2}^{2},}
\end{align}
where $\mathbf{\dot{a}}\left( \theta ,N \right)\triangleq \partial \mathbf{a}\left( \theta ,N \right)/\partial \theta$, $\mathbf{b}_{k}^{H}={{\mathbf{a}}^{H}}\left( {{\theta }_{k,\text{AOD}}},N \right){{\mathbf{\Theta }}_{k}}{{\mathbf{H}}_{k}}{{\mathbf{W}}_{k}}{{\mathbf{X}}_{k}}$ and $\mathbf{c}_{k}^{H}={{{\mathbf{\dot{a}}}}^{H}}\left( {{\theta }_{k,\text{AOD}}},N \right){{\mathbf{\Theta }}_{k}}{{\mathbf{H}}_{k}}{{\mathbf{X}}_{k}}$. Here, we omit the derivative calculation process, as it is similar to the process in the previous subsection.
\subsection{CRB for Location Estimation }\label{crb_location}
Since the location cannot be directly estimated from the signal, the Fisher Information Matrix (FIM) of the location can be derived using the chain rule, i.e.,
\begin{align}\label{fisherx}
\mathbf{F}\left( \mathbf{x} \right)=\frac{\partial \left[ \boldsymbol{\tau } ,\boldsymbol{\theta }  \right]}{\partial \mathbf{x}}\mathbf{F}\left( \boldsymbol{\tau } ,\boldsymbol{\theta }  \right){{\left( \frac{\partial \left[ \boldsymbol{\tau } ,\boldsymbol{\theta }  \right]}{\partial \mathbf{x}} \right)}^{T}}
\end{align}
where $\mathbf{F}\left( \boldsymbol{\tau } ,\boldsymbol{\theta } \right)=\text{diag}\left( \mathbf{F}\left( {{\boldsymbol\theta }_{\text{AOA}}} \right)+\mathbf{F}\left( {{\boldsymbol\theta }_{\text{AOD}}} \right),\mathbf{F}\left( \boldsymbol{\tau } \right) \right)$ and 
${\partial  \left[ \boldsymbol{\tau } ,\boldsymbol{\theta }  \right]}/{\partial {{\mathbf{x}}}}$ represents the geometric relationship, which can be expressed as 
\begin{align}
\frac{\partial \left[ \boldsymbol{\tau } ,\boldsymbol{\theta }  \right]}{\partial {{\mathbf{x}}}}=\left[ \begin{matrix}
   \frac{\partial \boldsymbol\tau }{\partial {{x}_{\text{T}}}} & \frac{\partial \boldsymbol\theta }{\partial {{x}_{\text{T}}}}  \\
   \frac{\partial \boldsymbol\tau }{\partial {{y}_{\text{T}}}} & \frac{\partial \boldsymbol\theta }{\partial {{y}_{\text{T}}}} \\
\end{matrix} \right],
\end{align}
where $\partial \boldsymbol\tau /\partial {{x}_{\text{T}}}=\left[ \cos {{\theta }_{1}}+\cos {{\theta }_{1}},\cdots ,\cos {{\theta }_{K}}+\cos {{\theta }_{K}} \right]/c \in {{\mathbb{R}}^{{{K}^{2}}}}$,
	$\partial \boldsymbol\tau /\partial {{y}_{\text{T}}}\ =\left[ \sin {{\theta }_{1}}+\sin {{\theta }_{1}},\cdots ,\sin {{\theta }_{K}}+\sin {{\theta }_{K}} \right]/c \in {{\mathbb{R}}^{{{K}^{2}}}}$,
$\partial \boldsymbol\theta /\partial {{x}_{\text{T}}}\ =-\left[ \sin {{\theta }_{1}}/{{d}_{1}}\ ,\cdots ,\sin {{\theta }_{K}}/{{d}_{K}}\  \right]\in {{\mathbb{R}}^{K}}$ and
$\partial \boldsymbol\theta /\partial {{y}_{\text{T}}}\ =\left[ \cos {{\theta }_{1}}/{{d}_{1}}\ ,\cdots ,\cos {{\theta }_{K}}/{{d}_{K}}\  \right]\in {{\mathbb{R}}^{K}}$.
Then, the CRB for the estimation of $\mathbf{x}$ can be expressed as
\begin{align}\label{crbx}
\text{CRB}\left( \mathbf{x} \right)=\text{tr}\left( {{\mathbf{F}}^{-1}}\left( \mathbf{x} \right) \right)
\end{align}

\section{Delay and Angle Estimation Algorithms}
In this section, we propose angle and cascade delay estimation algorithms.  The estimation of the cascade delay is relatively straightforward. For sensors on each semi-passive IRS, the received signal ${{\mathbf{R}}_{l}}$ is used to search for the time delay ${{\tau }_{l,k}}$ through the following matched filtering.
\begin{align}
{{\hat{\tau }}_{l,k}}=\underset{\tau }{\mathop{\arg \max }}\,{{\left\| {{\mathbf{R}}_{l}}\left( t \right)\mathbf{X}_{k}^{H}\left( t-\tau -{{\tau }_{\text{B2I,}k}} \right) \right\|}_{2}},
\end{align}
and ${{\mathbf{R}}_{l,k}}$ is obtained through the following matched filtering
\begin{align}\label{rlk}
&{{\mathbf{R}}_{l,k}} ={{\mathbf{R}}_{l}}\left( t \right)\mathbf{X}_{k}^{H}\left( t-{{{\hat{\tau }}}_{l,k}} \right). \nonumber\\
& ={{\alpha }_{l,k}}\mathbf{a}\left( {{\theta }_{l,\text{AOA}}},M \right){{\mathbf{a}}^{H}}\left( {{\theta }_{k,\text{AOD}}},N \right){{\mathbf{S}}_{k}}+{{\mathbf{N}}_{l}}\mathbf{X}_{k}^{H}\left( t-{{{\hat{\tau }}}_{l,k}}  \right),
\end{align}
where {\color{blue}${{\mathbf{S}}_{k}}={{\mathbf{\Theta }}_{k}}{{\mathbf{H}}_{k}}{{\mathbf{W}}_{k}}\left( {{\mathbf{X}}_{k}}\mathbf{X}_{k}^{H} \right)={{\mathbf{\Theta }}_{k}}{{\mathbf{H}}_{k}}{{\mathbf{W}}_{k}}$.}
Since path gain is related to distance and the target's location is unknown in the angle estimation problem, the path gain is unknown. As a result, the traditional angle estimation problem can be expressed as
\begin{align}
\underset{{{\theta }_{l,\text{AOA}}},{{\theta }_{k,\text{AOD}}},{{\alpha }_{l,k}}}{\mathop{\min }}\,\left\| {{\mathbf{R}}_{l,k}}-{{\alpha }_{l,k}}{{\mathbf{a}}}\left( {{\theta }_{l,\text{AOA}}} \right)\mathbf{a}^{H}\left( {{\theta }_{k,\text{AOD}}} \right){{\mathbf{S}}_{k}} \right\|_{2}^{2}.
\end{align}
Because of the non-convexity of the steering vector and the coupling between variables, this problem cannot be directly solved using convex optimization methods.

The traditional MUSIC, compressive sensing, and ESPRIT algorithms, due to their complexity, discretization, and sensitivity to noise, may not achieve optimal performance. Furthermore, these algorithms cannot simultaneously utilize observation samples from different angles for joint angle estimation, which limits the full potential of the collaborative localization system.

\subsection{Atomic Norm Minimization Formulation}
First, we introduce the concept of an atomic set.
A 2-D atomic norm set $\mathcal{A}$ can be represented as
\begin{align}
\mathcal{A}=\left\{ \mathbf{A}\left( {{\theta }_{l,\text{AOA}}},{{\theta }_{k\text{,AOD}}} \right)|{{\theta }_{l,\text{AOA}}},{{\theta }_{k\text{,AOD}}}\in \left[ {-\pi }/{2},{\pi}/{2}\right] \right\},
\end{align}
with $\mathbf{A}\left( {{\theta }_{l,\text{AOA}}},{{\theta }_{k\text{,AOD}}} \right)=\mathbf{a}\left( {{\theta }_{l,\text{AOA}}},M \right){{\mathbf{a}}^{H}}\left( {{\theta }_{k,\text{AOD}}},N \right)\in {{\mathbb{C}}^{M\times N}}$.
The atomic norm of a matrix $\mathbf{X}$ is defined as:
\begin{align}
{{\left\| \mathbf{X} \right\|}_{\mathcal{A}}}=\inf \left\{ \sum\limits_{l,k}{{{\alpha }_{l,k}}}|\mathbf{X}=\sum\limits_{l,k}{{{\beta }_{l,k}}\mathbf{A}\left( {{\theta }_{l,\text{AOA}}},{{\theta }_{k\text{,AOD}}} \right)}\right\}.
\end{align}
with ${{\beta }_{l,k}}>0$ represents the weight coefficient of each atomic term. By minimizing the atomic norm of $\mathbf{X}$, we can decompose $\mathbf{X}$ within the atomic set, thereby obtaining the steering vectors corresponding to the components. {\color{blue}It is noted that the signal model in (\ref{rlk}) does not directly fit the standard atomic normminimization framework, which typically decomposes a signal into a linear combination of atoms with scalar coefficients. To make the problem compatible with the ANM framework, a critical signal pre-processing step is performed first. Specifically, we eliminate the effect of the known matrix ${{\mathbf{S}}_{k}}$ by right-multiplying the received signal matrix ${{\mathbf{R}}_{l,k}}$ by the pseudo-inverse of ${{\mathbf{S}}_{k}}$, $\mathbf{S}_{k}^{\dagger }$. Let the pre-processed signal be $\mathbf{R}_{l,k}^{'}={{\mathbf{R}}_{l,k}}\mathbf{S}_{k}^{\dagger }\approx {{\alpha }_{l,k}}\mathbf{a}\left( {{\theta }_{l,\text{AOA}}},M \right){{\mathbf{a}}^{H}}\left( {{\theta }_{k,\text{AOD}}},N \right)$. However, due to the presence of noise, the received signal is often difficult to accurately decompose within the atomic set. Therefore, we consider minimizing the atomic norm within the $F$-norm domain of the received signal, i.e.,
\begin{align}
\text{(P1)}:~~~~ &\underset{\mathbf{X}}{\mathop{\min }}~~{{\left\| \mathbf{X} \right\|}_\mathcal{A}} \\
& ~\text{s.t.}~~~{{\left\| \mathbf{X}-{{\mathbf{R}}_{l,k}^{'}} \right\|}_{F}}\le \eta,
\end{align}}
where $\eta$  represents the allowable margin of error.
By utilizing the properties of the Vandermonde decomlocation, problem (P1) can be equivalently transformed into problem (P2), which can be expressed as
\begin{align}
\text{(P2)}:~~~ &\underset{\mathbf{X},\mathbf{Z},\mathbf{v},\mathbf{u}}{\mathop{\min }}~~~\text{tr}\left( \mathbf{T}\left( \mathbf{v} \right) \right)+\text{tr}\left( \mathbf{T}\left( \mathbf{u} \right) \right)\\
&~~\text{s.t.}~~~~~~{{\left\| \mathbf{X}-{{\mathbf{R}}_{l,k}^{'}} \right\|}_{F}}\le \eta ,\\
& ~~~~~~~~~~~\left[ \begin{matrix}
   \mathbf{T}\left( \mathbf{v} \right) & \mathbf{Z}  \\
   \mathbf{Z}^H &  \mathbf{T}\left( \mathbf{u} \right)  \\
\end{matrix} \right]\succeq 0,
\end{align}
where $\mathbf{T}\left( \mathbf{u} \right)$ denotes a Hermitian Toeplitz matrix with $\mathbf{u}$ as its main diagonal element \cite{yang2018sparse}. After the above equivalence, the problem becomes easier to solve. The previous subsection only introduced the basic one-dimensional atomic norm. In the subsequent solution process, we will apply the corresponding transformations to the one-dimensional atomic norm for the proposed model.

\subsection{Joint Angle Estimation Algorithm Based on ADMM}
In this subsection, we present a detailed description of the angle estimation algorithm. {\color{blue} First, we formulate a joint estimation framework where the objective function is a weighted sum over all $K$ collaborative paths, enabling the simultaneous processing of multiple observation matrices to estimate a common angle. This is a significant departure from traditional methods that handle a single observation matrix. Second, we propose a comprehensive multi-case framework, tailoring the signal processing and algorithm design based on the rank of the IRS-reflected signal matrix ${{\mathbf{S}}_{k}}$.} 

Specifically, for any angle ${{\theta }_{i}}$, there are $K$ observation samples containing information about ${{\theta }_{i,\text{AOA}}}$  (i.e., ${{\mathbf{R}}_{i,k}},\forall k$) and $K$ observation samples containing information about ${{\theta }_{i,\text{AOD}}}$  (i.e., ${{\mathbf{R}}_{l,i}},\forall l$). Joint estimation using all relevant observation samples achieves better performance compared to estimation based on single observation sample. Therefore, we first estimate ${{\theta }_{i,\text{AOA}}}$ and ${{\theta }_{i,\text{AOA}}}$ based on ${{\mathbf{R}}_{i,k}},\forall k$ and ${{\mathbf{R}}_{l,i}},\forall l$, respectively, and obtain ${{\theta }_{i}}$ through weighted combination of ${{\theta }_{i,\text{AOA}}}$ and ${{\theta }_{i,\text{AOD}}}$. Through the equivalence transformation in previous section, the estimation of ${{\theta }_{i,\text{AOA}}}$ can be expressed as
\begin{subequations}
\begin{alignat}{1}
\text{(P3)}  &\underset{{{\mathbf{v}}_{i}},\left\{ {{\mathbf{u}}_{k}} \right\},\left\{ {{\mathbf{Z}}_{i,k}} \right\}}{\mathop{\min }} \sum\limits_{k=1}^{K}{{{e}_{i,k}}\left[ \text{tr}\left( \mathbf{T}\left( {{\mathbf{u}}_{k}} \right)\right)+ \text{tr}\left(\mathbf{T}\left( {{\mathbf{v}}_{i}} \right) \right) \right.} \\
 &~~~~~~~~~~~~~~~\left. +\lambda \left\| {{\mathbf{Z}}_{i,k}}{{\mathbf{S}}_{k}}-{{\mathbf{R}}_{i,k}} \right\|_{2}^{2} \right] \nonumber\\ 
 &~~~~~~~\text{s.t.}~~~~\left[ \begin{matrix}
   \mathbf{T}\left( {{\mathbf{v}}_{i}} \right) & {{\mathbf{Z}}_{i,k}}  \\
   \mathbf{Z}_{i,k}^{H} & \mathbf{T}\left( {{\mathbf{u}}_{k}} \right) \\
\end{matrix} \right]\succeq 0,\forall k,
\end{alignat}
\end{subequations}
where ${{e}_{i,k}}=\left\| {{\mathbf{R}}_{i,k}} \right\|_{2}^{2}$ is a weighted coefficient representing the energy of the received signal, which characterizes the confidence level for each observation sample. Problem (P3) is non-convex due to 
$\left\| {{\mathbf{Z}}_{i,k}}{{\mathbf{S}}_{k}}-{{\mathbf{R}}_{i,k}} \right\|_{2}^{2}$ is an indefinite quadratic form and the semi-definite constraints. To address this, we choose to project the corresponding matrix onto the semi-definite cone and use ${{\mathbf{Z}}_{i,k}}-{{\mathbf{R}}_{i,k}}\mathbf{{S}}_{k}^{\dagger }$ to replace ${{\mathbf{Z}}_{i,k}}{{\mathbf{{S}}}_{k}}-{{\mathbf{R}}_{i,k}}$ for characterizing the 
reconstruction error of the signal. However, the Moore-Penrose inverse of $\mathbf{{S}}_{k}$ depends on its rank, which will affect the subsequent solution process. Therefore, we categorize the discussion based on the rank of $\mathbf{{S}}_{k}$ as follows:

\subsubsection{$\text{rank}\left( {{\mathbf{{S}}}_{k}}\right)=N$}
Since the rank of ${{\mathbf{{S}}}_{k}}$ is $N$, the  Moore-Penrose inverse of ${{\mathbf{{S}}}_{k}}$ can be denoted as 
$\mathbf{S}_{k}^{\dagger }=\mathbf{S}_{k}^{H}{{\left( {{\mathbf{S}}_{k}}\mathbf{S}_{k}^{H} \right)}^{-1}}$.
Then, the ${{\mathbf{R}}_{i,k}}$ can be multiplied on the right by ${{\mathbf{{S}}}_{k}}$ and $\left\| {{\mathbf{Z}}_{i,k}}-{{\mathbf{R}}_{i,k}}\mathbf{S}_{k}^{\dagger } \right\|_{2}^{2}$ is a convex function with respect to ${{\mathbf{Z}}_{i,k}}$. Then the Lagrangian function of (P3) can be formulated as (\ref{l1}), where ${{\mathbf{G}}_{i,k}}$ represents the Lagrange multiplier, and $\rho$ is the penalty factor. For ease of subsequent representation, the matrix ${{\mathbf{G}}_{i,k}},{{\mathbf{Z}}_{i,k}}$ is expressed as the following block matrix:
\begin{equation}
{{\mathbf{G}}_{i,k}}=\left[ \begin{matrix}
   {{\mathbf{G}}_{i,k,0}} & {{\mathbf{G}}_{i,k,1}}  \\
   \mathbf{G}_{i,k,1}^{H} & {{\mathbf{G}}_{i,k,2}}  \\
\end{matrix} \right],{{\mathbf{Z}}_{i,k}}=\left[ \begin{matrix}
   {{\mathbf{Z}}_{i,k,0}} & {{\mathbf{Z}}_{i,k,1}}  \\
   \mathbf{Z}_{i,k,1}^{H} & {{\mathbf{Z}}_{i,k,2}}  \\
\end{matrix} \right],
\end{equation}
where ${{\mathbf{G}}_{i,k,0}},{{\mathbf{Z}}_{i,k,0}}\in {{\mathbb{C}}^{M\times M}}$, ${{\mathbf{G}}_{i,k,1}},{{\mathbf{Z}}_{i,k,1}}\in {{\mathbb{C}}^{M\times N}}$ and ${{\mathbf{G}}_{i,k,2}},{{\mathbf{Z}}_{i,k,2}}\in {{\mathbb{C}}^{N\times N}}$.

\begin{figure*}[!t]
	\normalsize
\begin{align}\label{l1}
  & \mathcal{L}_{1}\left( {{\mathbf{v}}_{i}},{{\mathbf{u}}_{k}},{{\mathbf{X}}_{i,k}},{{\mathbf{G}}_{i,k}},{{\mathbf{Z}}_{i,k}} \right)=\sum\limits_{k=1}^{K}{{{e}_{i,k}}\left[ \text{tr}\left( \mathbf{T}\left( {{\mathbf{u}}_{k}} \right)+\mathbf{T}\left( {{\mathbf{v}}_{i}} \right) \right)+\lambda \left\| {{\mathbf{Z}}_{i,k}}-{{\mathbf{R}}_{i,k}}\mathbf{S}_{k}^{\dagger } \right\|_{2}^{2} \right.} \nonumber\\ 
 & \left. +\left\langle {{\mathbf{G}}_{i,k}},\left( {{\mathbf{X}}_{i,k}}-\left[ \begin{matrix}
   \mathbf{T}\left( {{\mathbf{v}}_{i}} \right) & {{\mathbf{Z}}_{i,k}}  \\
   \mathbf{Z}_{i,k}^{H} & \mathbf{T}\left( {{\mathbf{u}}_{k}} \right)  \\
\end{matrix} \right] \right) \right\rangle +\rho \left\| {{\mathbf{X}}_{i,k}}-\left( \begin{matrix}
   \mathbf{T}\left( {{\mathbf{v}}_{i}} \right) & {{\mathbf{Z}}_{i,k}}  \\
   \mathbf{Z}_{l,i}^{H} & \mathbf{T}\left( {{\mathbf{u}}_{k}} \right)  \\
\end{matrix} \right) \right\|_{2}^{2} \right],
\end{align}
	\hrulefill
	\vspace*{1pt}
\end{figure*}
These variables can be divided into three parts and updated sequentially by solving the corresponding subproblems. Specifically, the subproblems and
their solutions are elaborated as follows.
\begin{itemize}
\item Subproblem with respect to ${{\mathbf{v}}_{i}},{{\mathbf{u}}_{k}},{{\mathbf{X}}_{i,k}}$
\end{itemize}
By computing the derivative of $\mathcal{L}_{1}\left( {{\mathbf{v}}_{i}},{{\mathbf{u}}_{k}},{{\mathbf{X}}_{i,k}},{{\mathbf{Z}}_{i,k}},{{\mathbf{G}}_{i,k}} \right)$, we have
\begin{align}
 & {{\nabla }_{{{\mathbf{v}}_{i}}}}\mathcal{L}_{1}=\sum\limits_{k=1}^{K}{{{e}_{i,k}}\left\{ M{{\mathbf{e}}_{M,1}}-f\left( {{\mathbf{G}}_{i,k,0}} \right)\right.} \nonumber\\ 
 &~~~~~~~+\left. 2\rho \left[ f\left( {{\mathbf{Z}}_{i,k,0}} \right)-f\left( \mathbf{T}\left( {{\mathbf{u}}_{k}} \right) \right) \right] \right\},\\ 
   & {{\nabla }_{{{\mathbf{u}}_{k}}}}\mathcal{L}_{1}=N{{\mathbf{e}}_{N,1}}-f\left( {{\mathbf{G}}_{i,k,2}} \right)+2\rho \left[ f\left( \mathbf{T}\left( {{\mathbf{u}}_{k}} \right) \right)-f\left( {{\mathbf{Z}}_{i,k,2}} \right) \right],\\
 & {{\nabla }_{{{\mathbf{X}}_{i,k}}}}\mathcal{L}_{1}=\left( {{\mathbf{X}}_{i,k}}-{{\mathbf{R}}_{i,k}}\mathbf{S}_{k}^{\dagger } \right)-{{\mathbf{G}}_{i,k,1}}+\rho \left( 2{{\mathbf{X}}_{i,k}}-2{{\mathbf{Z}}_{i,k,1}} \right) \
\end{align}
where $f\left( {{\mathbf{X}}},N \right)$ is a mapping from a matrix $\mathbf{X}\in {{\mathbb{C}}^{N\times N}}$ to a vector, which is defined as 
\begin{align}
{{\left[ f\left( \mathbf{X},N \right) \right]}_{n}}=\left\{ \begin{aligned}
  & \sum\limits_{i=1}^{N}{{{\left[ \mathbf{X} \right]}_{i,i}}},n=1, \\ 
 & 2\sum\limits_{j-i=n}^{N}{{{\left[ \mathbf{X} \right]}_{i,j}}},n>1. \\ 
\end{aligned} \right.
\end{align}
By setting these derivatives to 0, ${{\mathbf{v}}_{i}},{{\mathbf{u}}_{k}},{{\mathbf{X}}_{i,k}}$ are updated by
\begin{align}\label{uk}
  & {{\left[ {{\mathbf{u}}_{k}} \right]}_{n}}=\frac{1}{g\left( n,N \right)}{{\left[ \frac{f\left( {{\mathbf{G}}_{i,k,2}} \right)+2\rho f\left( {{\mathbf{Z}}_{i,k,2}} \right)-N{{\mathbf{e}}_{N,1}}}{2\rho } \right]}_{n}},\\ 
 & {{\left[ {{\mathbf{v}}_{i}} \right]}_{m}}=\frac{1}{g\left( m,M \right)}  \nonumber\\ 
 & {{\left[ \frac{\sum\limits_{l=1}^{K}{{{e}_{i,k}}\left[ f\left( {{\mathbf{G}}_{i,k,0}},M \right)+2\rho f\left( {{\mathbf{Z}}_{i,k,0}},M \right)-M{{\mathbf{e}}_{M,1}} \right]}}{\sum\limits_{l=1}^{K}{{{e}_{i,k}}}2\rho } \right]}_{m}},\\ 
 & {{\mathbf{X}}_{i,k}}=\left( \lambda {{\mathbf{R}}_{i,k}}\mathbf{S}_{k}^{\dagger }+{{\mathbf{G}}_{i,k,1}}+2\rho {{\mathbf{Z}}_{i,k,1}} \right){{\left( \lambda +2\rho  \right)}^{-1}},
\end{align}
where $t$ is the iteration index and function $g\left( \cdot  \right)$ is defined as
\begin{align}
g\left( n,N \right)=\left\{ \begin{aligned}
  & N,n=1, \\ 
 & 2\left( N-n+1 \right),n>1, \\ 
\end{aligned} \right.
\end{align}

\begin{itemize}
\item Subproblem with respect to ${{\mathbf{Z}}_{i,k}}$
\end{itemize}
The update of ${{\mathbf{Z}}_{i,k}}$ can be expressed as solving the following problem:
\begin{align}
\mathbf{Z}_{i,k}^{r}=\underset{{{\mathbf{Z}}_{i,k}}\ge 0}{\mathop{\arg \min }}\,{{\mathcal{L}}_{1}}\left( \mathbf{v}_{i}^{r},\mathbf{u}_{k}^{r},\mathbf{X}_{i,k}^{r},{{\mathbf{Z}}_{i,k}^{r}},\mathbf{G}_{i,k}^{r-1}. \right)
\end{align}
Because the positive semi-definite constraint is a non-convex constraint, we first ignore this constraint to solve the problem, leading to the following result
\begin{align}
\mathbf{\tilde{Z}}_{i,k}^{r}=\left[ \begin{matrix}
   \mathbf{T}\left( \mathbf{v}_{i}^{r} \right) & \mathbf{X}_{i,k}^{r}  \\
   {{\left( \mathbf{X}_{i,k}^{r} \right)}^{H}} & \mathbf{T}\left( \mathbf{u}_{k}^{r} \right)  \\
\end{matrix} \right]-\frac{\mathbf{G}_{i,k}^{r-1}}{2\rho }.
\end{align}
To satisfy the positive semi-definite constraint, we perform eigenvalue decomlocation on $\mathbf{\tilde{Z}}_{i,k}^{r+1}$, resulting in 
$\mathbf{\tilde{Z}}_{i,k}^{r}=\sum\nolimits_{q=1}^{{{R}_{i,k}}}{{{\lambda }_{q}}{{\mathbf{v}}_{q}}\mathbf{v}_{q}^{H}}$, where ${{R}_{i,k}}$ is the rank of $\mathbf{\tilde{Z}}_{i,k}^{r}$ and
\begin{align}
\mathbf{Z}_{i,k}^{r}=\sum\nolimits_{q=1}^{{{R}_{i,k}}}{\max \left( {{\lambda }_{q}},0 \right){{\mathbf{v}}_{q}}\mathbf{v}_{q}^{H}}.
\end{align}
\begin{itemize}
\item Subproblem with respect to ${{\mathbf{G}}_{i,k}}$
\end{itemize}
The update for ${{\mathbf{G}}_{i,k}}$ is based on the multiplier update rule, i.e.,
\begin{align}
\mathbf{G}_{i,k}^{r}=\mathbf{G}_{i,k}^{r-1}+\rho \left( {{\mathbf{Z}}_{i,k}^{r}}-\left[ \begin{matrix}
   \mathbf{T}\left( \mathbf{v}_{i}^{r} \right) & \mathbf{X}_{i,k}^{r}  \\
   {{\left( \mathbf{X}_{i,k}^{r} \right)}^{H}} & \mathbf{T}\left( \mathbf{u}_{k}^{r} \right)  \\
\end{matrix} \right] \right).
\end{align}
At this point, we have derived the closed-form solutions for updating each variable. By alternately updating these variables until convergence, the entire algorithm is summarized in \textbf{Algorithm 1}.
\begin{algorithm}[t]
	\caption{Proposed Alternating Optimization Algorithm.} 
	\begin{algorithmic}[1]
		\State$\textbf{Input}$: Initialize ${{\rho }}$,  ${\lambda}$, $\mathbf{u}_{k}^{0},\mathbf{v}_{i}^{0},\mathbf{X}_{i,k}^{0},\mathbf{Z}_{i,k}^{0},\mathbf{G}_{i,k}^{0}$, threshold ${{\varepsilon }_{1}}$, ${{\varepsilon }_{2}}$ and iteration index $r=1$ 
		\Repeat:
        \State Update  $\mathbf{u}_{k}^{r},\mathbf{v}_{i}^{r},\mathbf{X}_{i,k}^{r}$ according to (34), (35) and (36), respectively. 
        \State Update $\mathbf{Z}_{i,k}^{r}$ by solving according to (40).
        \State Update $\mathbf{G}_{i,k}^{r}$ by solving according to (41).
        \State $r=r+1$.
\Until 
$\left\| \mathbf{X}_{i,k}^{r}-\mathbf{X}_{i,k}^{r-1} \right\|_{2}^{2}\le {{\varepsilon }_{1}}$ and $\left\| \mathbf{X}_{i,k}^{r}-\mathbf{X}_{i,k}^{r-1} \right\|_{2}^{2}\le {{\varepsilon }_{2}}$.
\State$\textbf{Output}$: $\mathbf{v}_{i}$
	\end{algorithmic}
\end{algorithm}

After solving for  $\mathbf{T}\left( {{\mathbf{v}}_{i}} \right)$, we perform eigenvalue decomlocation, and the normalised eigenvector corresponding to the largest eigenvalue is taken as the estimation of $a\left( {{\theta }_{i,\text{AOA}}},M \right)$. The estimation of ${\theta }_{i,\text{AOD}}$ is similar to that of ${\theta }_{i,\text{AOA}}$. Specifically, replace Lagrange function $\mathcal{L}_{1}\left( {{\mathbf{v}}_{i}},{{\mathbf{u}}_{k}},{{\mathbf{X}}_{i,k}},{{\mathbf{G}}_{i,k}},{{\mathbf{Z}}_{i,k}} \right)$ with $\mathcal{L}_{1}\left( {{\mathbf{v}}_{k}},{{\mathbf{u}}_{i}},{{\mathbf{X}}_{k,i}},{{\mathbf{G}}_{k,i}},{{\mathbf{Z}}_{k,i}} \right)$ and the subsequent variable update process can follow \textbf{Algorithm 1}.

\subsubsection{$1<\text{rank}\left( {{\mathbf{{S}}}_{k}} \right)<N$} The  singular value decomlocation (SVD) of matrix ${{\mathbf{S}}_{k}}={{\mathbf{U}}_{k}}{{\Sigma }_{k}}\mathbf{V}_{k}^{H}$.
Based on the result of the singular value decomlocation and substituting it into (\ref{rlk}), ${{\mathbf{R}}_{i,k}}$ can be expressed as
\begin{align}
{{\mathbf{R}}_{i,k}}={{\alpha }_{i,k}}\mathbf{a}\left( {{\theta }_{i,\text{AOA}}},M \right){{\mathbf{a}}^{H}}\left( {{\theta }_{k,\text{AOD}}},N \right){{\mathbf{U}}_{k}}{{\mathbf{\Sigma }}_{k}}\mathbf{V}_{k}^{H}+{{\mathbf{N}}_{i,k}}
\end{align}
where ${{\mathbf{U}}_{k}},\in {{\mathbb{C}}^{N\times rank\left( {{\mathbf{S}}_{k}} \right)}},{{\mathbf{\Sigma }}_{k}}\in {{\mathbb{C}}^{rank\left( {{\mathbf{S}}_{k}} \right)\times rank\left( {{\mathbf{S}}_{k}} \right)}},{{\mathbf{V}}_{k}}\in {{\mathbb{C}}^{L\times rank\left( {{\mathbf{S}}_{k}} \right)}}$, ${{\mathbf{N}}_{i,k}}={{\mathbf{N}}_{i}}\mathbf{X}_{k}^{H}\left( t-{{{\hat{\tau }}}_{i,k}} \right)$ represents the Gaussian white noise after matched filtering.
It is noted that when the vector ${{\mathbf{a}}^{H}}\left( {{\theta }_{k,\text{AOD}}},N \right)$ containing the information of ${{\theta }_{k,\text{AOD}}}$ is multiplied by the left singular matrix ${{\mathbf{U}}_{k}}$, and ${{\mathbf{a}}^{H}}\left( {{\theta }_{k,\text{AOD}}},N \right)$ is compressed into a lower-dimensional vector, which leads to the destruction of the structure of the steering vector. Therefore, we reconstruct the atomic set as
\begin{align}
\mathcal{A}=\left\{ \mathbf{A}\left( {{\theta }_{l,\text{AOA}}},\mathbf{b} \right)\in {{\mathbb{C}}^{M\times rank\left( {{\mathbf{S}}_{k}} \right)}}|{{\theta }_{l,\text{AOA}}}\in \left[ -\pi /2\ ,\pi /2 \right] \right\},
\end{align}
with $\mathbf{A}\left( {{\theta }_{l,\text{AOA}}},\mathbf{b} \right)=\mathbf{a}\left( {{\theta }_{l,\text{AOA}}},M \right){{\mathbf{b}}^{H}}$ and ${{\left\| \mathbf{b} \right\|}_{2}}=1$. By multiplying matrix ${{\left( {{\mathbf{\Sigma }}_{k}}{{\mathbf{V}}_{k}} \right)}^{\dagger }}$ on the right, the received signal ${{\mathbf{R}}_{i,k}}$ can be re-expressed as the sum of the atoms and noise, i.e., 
\begin{align}
{{\mathbf{R}}_{i,k}}{{\left( {{\mathbf{\Sigma }}_{k}}{{\mathbf{V}}_{k}} \right)}^{\dagger }}={{\beta }_{i,k}}\mathbf{A}\left( {{\theta }_{l,\text{AOA}}},\mathbf{b} \right)+{{\mathbf{N}}_{i,k}}{{\left( {{\mathbf{\Sigma }}_{k}}{{\mathbf{V}}_{k}} \right)}^{\dagger }},
\end{align}
where ${{\beta }_{i,k}}{{\mathbf{b}}^{H}}$ equivalently represents ${{\alpha }_{i,k}}{{\mathbf{a}}^{H}}\left( {{\theta }_{k,\text{AOD}}},N \right){{\mathbf{U}}_{k}}$ in ${{\mathbf{R}}_{i,k}}$, and ${{\beta }_{i,k}}{{\mathbf{b}}^{H}}$ is more consistent with the definition of atomic norm compared to ${{\alpha }_{i,k}}{{\mathbf{a}}^{H}}\left( {{\theta }_{k,\text{AOD}}},N \right){{\mathbf{U}}_{k}}$. Then, we compute the atomic norm of matrix ${{\mathbf{R}}_{i,k}}{{\left( {{\mathbf{\Sigma }}_{k}}{{\mathbf{V}}_{k}} \right)}^{\dagger }}$ using a method similar to \textbf{Algorithm 1}, and the corresponding Lagrangian function is modified as shown in (\ref{l2}).
Specifically, these variables can be divided into three parts 
(${{\mathbf{v}}_{i}},{{\mathbf{P}}_{k}},{{\mathbf{X}}_{i,k}}$), ${{\mathbf{G}}_{i,k}}$ and ${{\mathbf{Z}}_{i,k}}$ and updated sequentially. Since the specific update method is similar to \textbf{Algorithm 1}.
\begin{figure*}[!t]
	\normalsize
\begin{align}\label{l2}
  & {\mathcal{L}_{2}}\left( {{\mathbf{v}}_{i}},{{\mathbf{P}}_{k}},{{\mathbf{X}}_{i,k}},{{\mathbf{G}}_{i,k}},{{\mathbf{Z}}_{i,k}} \right)=\sum\limits_{k=1}^{K}{{{e}_{i,k}}\left[ \text{tr}\left( \mathbf{T}\left( {{\mathbf{v}}_{i}} \right) \right)+\text{tr}\left( {{\mathbf{P}}_{k}} \right) \right.}+\lambda \left\| {{\mathbf{Z}}_{i,k}}-{{\mathbf{R}}_{i,k}}{{\mathbf{S}}_{k}}^{\dagger } \right\|_{2}^{2} \nonumber\\ 
 & \left. \left\langle {{\mathbf{G}}_{i,k}},\left( {{\mathbf{X}}_{i,k}}-\left[ \begin{matrix}
   \mathbf{T}\left( {{\mathbf{v}}_{i}} \right) & {{\mathbf{Z}}_{i,k}}  \\
   \mathbf{Z}_{i,k}^{H} & {{\mathbf{P}}_{k}}  \\
\end{matrix} \right] \right) \right\rangle +\rho \left\| {{\mathbf{X}}_{i,k}}-\left[ \begin{matrix}
   \mathbf{T}\left( {{\mathbf{v}}_{i}} \right) & {{\mathbf{Z}}_{i,k}}  \\
   \mathbf{Z}_{i,k}^{H} & {{\mathbf{P}}_{k}}  \\
\end{matrix} \right] \right\|_{2}^{2} \right], 
\end{align}
\hrulefill
\vspace*{1pt}
\end{figure*}
Once the estimation of angle ${{\theta }_{l,\text{AOA}}}$ is completed, the next step is to estimate angle ${{\theta }_{k,\text{AOD}}}$. The corresponding problem can be expressed as 
\begin{align}
\underset{{{\alpha }_{i,k}},{{\theta }_{k,\text{AOD}}}}{\mathop{\min }}\,\left\| {{\alpha }_{i,k}}\mathbf{\hat{a}}\left( {{\theta }_{i,\text{AOA}}},M \right){{\mathbf{a}}^{H}}\left( {{\theta }_{k,\text{AOD}}},N \right){{\mathbf{S}}_{k}}-{{\mathbf{R}}_{l,k}} \right\|_{2}^{2}.
\end{align}
To simplify the problem, by performing the left multiplication of vector ${{\mathbf{\hat{a}}}^{\dagger }}\left( {{\theta }_{i,\text{AOA}}},M \right)$ by matrix ${{\mathbf{R}}_{l,k}}$, the problem can be simplified to
\begin{align}
\underset{{{\alpha }_{i,k}},{{\theta }_{k,\text{AOD}}}}{\mathop{\min }}\,h\left( {{\alpha }_{i,k}},{{\theta }_{k,\text{AOD}}} \right)=\left\| {{\alpha }_{i,k}}{{\mathbf{S}}_{k}}\mathbf{a}\left( {{\theta }_{k,\text{AOD}}},N \right)-{{\mathbf{r}}_{i,k}} \right\|_{2}^{2},
\end{align}
where $\mathbf{r}_{i,k}^{H}={{\mathbf{\hat{a}}}^{\dagger }}\left( {{\theta }_{i,\text{AOA}}},M \right){{\mathbf{R}}_{i,k}}$. Due to the non-convexity of $\mathbf{a}\left( {{\theta }_{k,\text{AOD}}},N \right)$, we employ a gradient-based trust-region algorithm for estimation. Specifically, we use a second-order Taylor expansion $\tilde{h}$ to approximate $h$, i.e., 
\begin{align}
& \tilde{h}\left( {{\mathbf{y}}^{r}} \right)=h\left( {{\mathbf{y}}^{r-1}} \right)+{{\left( {{\mathbf{g}}^{r}} \right)}^{H}}\left( {{\mathbf{y}}^{r}}-{{\mathbf{y}}^{r-1}} \right) \nonumber\\ 
& +{{\left( {{\mathbf{y}}^{r}}-{{\mathbf{y}}^{r-1}} \right)}^{H}}{{\mathbf{G}}^{r}}\left( {{\mathbf{y}}^{r}}-{{\mathbf{y}}^{r-1}} \right),
\end{align}
where ${{\mathbf{y}}^{r}}=\left[ \alpha _{i,k}^{r},\theta _{k,\text{AOD}}^{r} \right]$, ${{\mathbf{g}}^{r}}$ and
${{\mathbf{G}}^{r}}$
represent the value of the optimization variable, gradient and Hessian matrix at the $r$-th iteration, respectively.
For the $r$-th iteration, the subproblem to be solved can be expressed as
\begin{align}
\text{(P4)}~~\underset{{{\mathbf{y}}^{r}}}{\mathop{\min }}\,\tilde{h}\left( {{\mathbf{y}}^{r}} \right)\text{   s}\text{.t}\text{.  }\left\| {{\mathbf{y}}^{r}}-{{\mathbf{y}}^{r-1}} \right\|_{2}^{2}\le {{l}^{r}},
\end{align}
where ${{l}^{r}}$ represents the upper bound of the search step size at the $r$-th iteration. This problem is a typical and common quadratic optimization problem, and here we omit the detailed solution process.
After solving the subproblem, the approximation degree of the Taylor expansion ${{s}^{r}}$ is computed to determine the range of the search step size upper bound for the next iteration, which can be expressed as
\begin{align}
{{s}^{r}}=\frac{h\left( {{\mathbf{y}}^{r}} \right)-h\left( {{\mathbf{y}}^{r-1}} \right)}{\tilde{h}\left( {{\mathbf{y}}^{r}} \right)-h\left( {{\mathbf{y}}^{r-1}} \right)}.
\end{align}
The updating and solving process is repeated until convergence, and the specific algorithm flow is shown in \textbf{Algorithm 2}.

\begin{algorithm}[t]
	\caption{Proposed AOD estimation Algorithm.} 
	\begin{algorithmic}[1]
		\State$\textbf{Input}$: Initialize ${{\mathbf{y}}^{0}}$,  ${{l}^{1}}$, threshold ${{\varepsilon }_{3}}$ and iteration index $r=1$ for outer loop
		\Repeat: inner loop
        \State Update ${{\mathbf{y}}^{r}}$ by solving (P4). 
        \State Update ${{\mathbf{s}}^{r}}$ according to (47). 
\If{${{\mathbf{s}}^{r}}>0.75$}          
    \State ${{l}^{r}} = 2{{l}^{r}}$.
\Else
    \State ${{l}^{r}}=\frac{1}{4}\left\| {{\mathbf{y}}^{r}}-{{\mathbf{y}}^{r-1}} \right\|_{2}^{2}$.
\EndIf
\State $r=r+1$.  
\Until 
$\left| h\left( {{\mathbf{y}}^{r}} \right)-h\left( {{\mathbf{y}}^{r-1}} \right) \right|/h\left( {{\mathbf{y}}^{r-1}} \right)\ \le {{\varepsilon }_{3}}$.
\State$\textbf{Output}$: ${{\hat{\theta }}_{k,\text{AOD}}}={{\left[ {{\mathbf{y}}^{r}} \right]}_{2}}$
	\end{algorithmic}
\end{algorithm}

 \subsubsection{$\text{rank}\left( {{\mathbf{{S}}}_{k}} \right)=1$}  The  singular value decomlocation (SVD) of matrix $ {{\mathbf{S}}_{k}}$ is given by
 ${{\mathbf{S}}_{k}}={{\mathbf{u}}_{k}}{{\Sigma }_{k}}\mathbf{v}_{k}^{H}$,
Based on the result of the singular value decomlocation and substituting it into (\ref{rlk}), ${{\mathbf{R}}_{i,k}}$ can be expressed as
\begin{align}
{{\mathbf{R}}_{i,k}}={{\alpha }_{i,k}}\mathbf{a}\left( {{\theta }_{i,\text{AOA}}},M \right){{\mathbf{a}}^{H}}\left( {{\theta }_{k,\text{AOD}}},N \right){{\mathbf{u}}_{k}}{{\sum }_{k}}\mathbf{v}_{k}^{H}+{{\mathbf{N}}_{i,k}}
\end{align}
where ${{\mathbf{N}}_{i,k}}={{\mathbf{N}}_{i}}\mathbf{X}_{k}^{H}\left( t-{{{\hat{\tau }}}_{i,k}} \right)$ represents the Gaussian white noise after matched filtering. 

It is noted that when the vector ${{\mathbf{a}}^{H}}\left( {{\theta }_{k,\text{AOD}}},N \right)$ containing the information of ${{\theta }_{k,\text{AOD}}}$ is multiplied by the left singular vector ${{\mathbf{u}}_{k}}$, the angular information is compressed into a scalar. At the same time, ${{\alpha }_{l,k}}$ is unknown, which results in the coupling of ${{\alpha }_{l,k}}$ and ${{\mathbf{a}}^{H}}\left( {{\theta }_{k,\text{AOD}}},N \right){{\mathbf{u}}_{k}}$, making ${{\theta }_{k,\text{AOD}}}$ unestimable.
Therefore, we use the one-dimensional ADMM to estimate \( a \), and the corresponding Lagrangian function is modified as shown in (\ref{l3}).

\begin{figure*}[!t]
	\normalsize
\begin{align}\label{l3}
  & {\mathcal{L}_{3}}\left( {{\mathbf{v}}_{i}},{{p}_{k}},{{\mathbf{X}}_{i,k}},{{\mathbf{G}}_{i,k}},{{\mathbf{z}}_{i,k}} \right)=\sum\limits_{k=1}^{K}{{{e}_{i,k}}\left[ \text{tr}\left( \mathbf{T}\left( {{\mathbf{v}}_{i}} \right) \right)+{{p}_{k}} \right.}+\lambda \left\| {{\mathbf{z}}_{i,k}}-{{\mathbf{R}}_{i,k}}{{\mathbf{S}}_{k}}^{\dagger } \right\|_{2}^{2} \nonumber\\ 
 & \left. \left\langle {{\mathbf{G}}_{i,k}},\left( {{\mathbf{X}}_{i,k}}-\left[ \begin{matrix}
   \mathbf{T}\left( {{\mathbf{v}}_{i}} \right) & {{\mathbf{z}}_{i,k}}  \\
   \mathbf{z}_{i,k}^{H} & {{p}_{k}}  \\
\end{matrix} \right] \right) \right\rangle +\rho \left\| {{\mathbf{X}}_{i,k}}-\left[ \begin{matrix}
   \mathbf{T}\left( {{\mathbf{v}}_{i}} \right) & {{\mathbf{z}}_{i,k}}  \\
   \mathbf{z}_{i,k}^{H} & {{p}_{k}}  \\
\end{matrix} \right] \right\|_{2}^{2} \right], 
\end{align}
\hrulefill
\vspace*{1pt}
\end{figure*}

\section{Location Estimation Algorithms}
In the previous section, the estimation of all time delay information $\boldsymbol{\tau }=\left[ {{\tau }_{1,1}},\cdots,{{\tau }_{K,K}} \right]\in {{\mathbb{R}}^{{{K}^{2}}\times 1}}$ and angle information $\boldsymbol\theta =\left[ {{\theta }_{1}},\cdots,{{\theta }_{K}} \right]\in {{\mathbb{R}}^{K\times 1}}$ has been completed.
\begin{figure}
\centerline{\includegraphics[width=8cm]{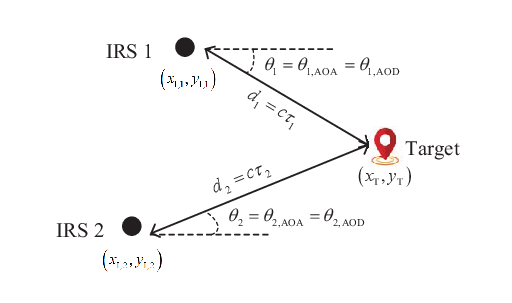}}
	\caption{The geometric relationship between delay, angle, and location.}
	\label{Fig1}
\end{figure}
Based on the geometric relationship shown in Fig. 2, the relationship between the time delay, angle, and the coordinates to be estimated can be expressed as
{\color{blue}
\begin{align}
  & -2{{x}_{\text{I},k}}{{x}_{\text{T}}}-2{{y}_{\text{I},k}}{{y}_{\text{T}}}+x_{\text{T}}^{2}+y_{\text{T}}^{2}=\tau_{k}^{2}c^{2}-x_{\text{I},k}^{2}-y_{\text{I},k}^{2},\forall k,\label{tAOAxb}\\ 
 & {{x}_{\text{T}}}\tan{{\theta }_{k}}-{{y}_{\text{T}}}={{x}_{\text{I},k}}\tan{{\theta }_{k}}-{{y}_{\text{I},k}},\forall k. \label{thetaaxb}
\end{align}}
We need to estimate $\left( {{x}_{\text{T}}},{{y}_{\text{T}}} \right)$ based on the cascade delay and angle estimates $\hat{\boldsymbol{\tau }}$
and $\hat{\boldsymbol{\theta }}$ according to (\ref{tAOAxb}) and (\ref{thetaaxb}).
The estimation of $\left( {{x}_{\text{T}}},{{y}_{\text{T}}} \right)$ faces three challenges: first, the directly estimated data is the cascade delay rather than the segment delay while segment delays are needed for localization. Then (\ref{tAOAxb}) contains quadratic information (i.e., $x_{\text{T}}^{2}+y_{\text{T}}^{2}$) about the location, and second, There exists estimation error in the coefficient matrix. 

To address the above challenges, we propose a three-stage algorithm. In the first stage, we estimate the segment delays based on the cascade delay. Then, in the second stage, we ignore the information in the quadratic terms and use weighted least squares to process error in the coefficient matrix to obtain a rough estimate. Finally, we use the information in the quadratic terms to refine the rough estimate.

\subsection{Stage 1} 
First, the segment delays ${{\tau }_{k}}$ and ${{\tau }_{l}}$ can be estimated based on the cascade delay estimate ${{\tau }_{l,k}}$. {\color{blue}Specifically, the error in the delay estimate ${{\tau }_{l,k}}$ can be modeled as ${{\hat{\tau }}_{l,k}}={{\tau }_{l,k}}+{{\varepsilon }_{l,k}}$, where ${{\tau }}_{l,k}$, ${{\hat{\tau }}_{l,k}}$, and ${{\varepsilon }_{l,k}}$ represent the true value, the true estimated value, and the estimation error of the cascade delay, respectively. }
Then, we construct the following equation to perform a weighted least squares estimation of ${{\tau }_{k}}$.
\begin{align}
\underbrace{\left[ \begin{matrix}
   2 & 0 & \cdots  & 0  \\
   1 & 1 & \cdots  & 0  \\
   \cdots  & \cdots  & \cdots  & \cdots   \\
   0 & 0 & \cdots  & 2  \\
\end{matrix} \right]}_{{{\mathbf{A}}_{1}}}\underbrace{\left[ \begin{matrix}
   {{\tau }_{1}}  \\
   {{\tau }_{2}}  \\
   \cdots   \\
   {{\tau }_{K}}  \\
\end{matrix} \right]}_{{{\mathbf{x}}_{1}}}=\underbrace{\left[ \begin{matrix}
   {{{\hat{\tau }}}_{1,1}}  \\
   {{{\hat{\tau }}}_{1,2}}  \\
   \cdots   \\
   {{{\hat{\tau }}}_{K,K}}  \\
\end{matrix} \right]}_{{{\mathbf{b}}_{1}}}+\underbrace{\left[ \begin{matrix}
   {{\varepsilon }_{1,1}}  \\
   {{\varepsilon }_{1,2}}  \\
   \cdots   \\
   {{\varepsilon }_{K,K}}  \\
\end{matrix} \right]}_{\Delta {{\mathbf{b}}_{1}}}.
\end{align}
Thus, the estimate of ${{\mathbf{x}}_{1}}$ can be expressed as
\begin{align}\label{esx1}
{{\mathbf{\hat{x}}}_{1}}={{\left( \mathbf{A}_{1}^{T}\mathbf{Q}_{1}^{-1}{{\mathbf{A}}_{1}} \right)}^{-1}}\mathbf{A}_{1}^{T}\mathbf{Q}_{1}^{-1}{{\mathbf{b}}_{1}},
\end{align}
Where $\mathbf{Q}_{1}$ is the autocorrelation matrix of the estimation error $\mathbf{b}_{1}$, which can be approximated through statistical averaging.

\subsection{Stage 2} 
The relationship between the estimated and true values of the segment delays and angles can be modeled as
\begin{align}
  & {{\tau}_{k}}={{{\hat{\tau}}}_{k}}+{{\varepsilon }_{\tau,k}},\forall k, \label{taoe}\\ 
 & {{\theta }_{k}}={{{\hat{\theta }}}_{k}}+{{\varepsilon }_{\theta,k}},\forall k,\label{thetae}
\end{align}
where ${{\varepsilon }_{k}}$ and ${{\varepsilon }_{k}}$ represent the estimation errors of the distance for the $k$-th path and the angle for the $k$-th IRS, respectively. By substituting equations (\ref{taoe}) and (\ref{thetae}) into equations (\ref{tAOAxb}) and (\ref{thetaaxb}) and defining the variables in equation (\ref{definationaxb}), we can establish the following system of equations:
\begin{align}
\underbrace{\left[ \begin{matrix}
   {{\mathbf{A}}_{\tau }}  \\
   {{\mathbf{A}}_{\theta }}+\Delta {{\mathbf{A}}_{\theta }}  \\
\end{matrix} \right]}_{{{\mathbf{A}}_{\mathbf{2}}}}\underbrace{\left[ \begin{aligned}
  & {{x}_{\text{T}}} \\ 
 & {{y}_{\text{T}}} \\ 
 & x_{\text{T}}^{2}+y_{\text{T}}^{2} \\ 
\end{aligned} \right]}_{{{\mathbf{x}}_{2}}}=\underbrace{\left[ \begin{matrix}
   {{\mathbf{b}}_{\tau }}  \\
   {{\mathbf{b}}_{\theta }}  \\
\end{matrix} \right]}_{{{\mathbf{b}}_{\mathbf{2}}}}+\underbrace{\left[ \begin{matrix}
   \Delta {{\mathbf{b}}_{\tau }}  \\
   \Delta {{\mathbf{b}}_{\theta }}  \\
\end{matrix} \right]}_{\Delta {{\mathbf{b}}_{\mathbf{2}}}},
\end{align}
where $\Delta {{\mathbf{b}}_{\tau }}\sim\mathcal{C}\mathcal{N}\left( {{\mathbf{0}}_{{{K}}}},\text{diag}\left( {{c}^{2}}\hat{\tau }_{1}^{2}\sigma _{1}^{2},\cdots ,{{c}^{2}}\hat{\tau }_{K}^{2}\sigma _{K}^{2} \right) \right)$, $\Delta {{\mathbf{b}}_{\theta }}\sim\mathcal{C}\mathcal{N}\left( {{\mathbf{0}}_{K}},\text{diag}\left( x_{\text{I},1}^{2}\sigma _{1}^{2}{{\sec }^{4}}{{{\hat{\theta }}}_{1}},\cdots ,x_{\text{I},K}^{2}\sigma _{K}^{2}{{\sec }^{4}}{{{\hat{\theta }}}_{K}} \right) \right)$ and $\text{vec}\left( \Delta {{\mathbf{A}}_{\theta }} \right)\sim\mathcal{C}\mathcal{N}\left( {{\mathbf{0}}_{3K}} \right.,\text{diag}\left( x_{\text{I},1}^{2}\sigma _{1}^{2}{{\sec }^{4}}{{{\hat{\theta }}}_{1}},\cdots , \right.$ $\left. \left. ,x_{\text{I},K}^{2}\sigma _{K}^{2}{{\sec }^{4}}{{{\hat{\theta }}}_{K}},{{\mathbf{0}}_{2K}} \right) \right)$.

First, we perform a variable substitution by defining $s\triangleq x_{\text{T}}^{2}+y_{\text{T}}^{2}$ and treating $s$ as a new variable independent of $x_{\text{T}}$ and $y_{\text{T}}$, so that the unknown terms become linear and the least squares estimation for $\mathbf{x}_2$ can be modeled as
\begin{align}
\text{(P4)}
& \underset{\Delta {{\mathbf{a}}_{\theta }},\Delta {{\mathbf{b}}_{\theta }},\Delta {{\mathbf{b}}_{\tau }}}{\mathop{\min }}\,\Delta \mathbf{a}_{\theta }^{T}\mathbf{R}_{\Delta {{\mathbf{a}}_{\theta }}}^{-1}\Delta {{\mathbf{a}}_{\theta }}+\Delta \mathbf{b}_{\theta }^{T}\mathbf{R}_{\Delta {{\mathbf{b}}_{\theta }}}^{-1}\Delta {{\mathbf{b}}_{\theta }} \nonumber\\
&~~~~~~~~~~~~~~~+
\Delta \mathbf{b}_{\tau }^{T}\mathbf{R}_{\Delta {{\mathbf{b}}_{\tau }}}^{-1}\Delta {{\mathbf{b}}_{\tau }} \nonumber\\ 
 & ~~~~~~\text{s.t.}~~~~~~{{\mathbf{A}}_{2}}{{\mathbf{x}}_{2}}={{\mathbf{b}}_{2}}+\Delta {{\mathbf{b}}_{2}}
\end{align}
The Lagrange function for P4 can be expressed as
\begin{align}
  & {\mathcal{L}_{4}}\left( \boldsymbol{\lambda }_{\theta }^{T},\boldsymbol{\lambda }_{\tau }^{T},\Delta {{\mathbf{a}}_{\theta }},\Delta {{\mathbf{b}}_{\theta }},\Delta {{\mathbf{b}}_{\tau }},\mathbf{x}_2 \right)=\boldsymbol{\lambda }_{\tau }^{T}\left( {{\mathbf{A}}_{\tau }}\mathbf{x}_2-{{\mathbf{b}}_{\tau }}-\Delta {{\mathbf{b}}_{\tau }} \right) \nonumber\\ 
 & +\boldsymbol{\lambda }_{\theta }^{T}\left( \left( {{\mathbf{A}}_{\theta }}+\left[ \begin{matrix}
   \Delta {{\mathbf{a}}_{\theta }} & {{\mathbf{0}}_{K\times 2}}  \\
\end{matrix} \right] \right)\mathbf{x}_2-{{\mathbf{b}}_{\theta }}-\Delta {{\mathbf{b}}_{\theta }} \right) \\ 
 & +\Delta \mathbf{a}_{\theta }^{T}\mathbf{R}_{\Delta {{\mathbf{A}}_{\theta }}}^{-1}\Delta {{\mathbf{a}}_{\theta }}+\Delta \mathbf{b}_{\theta }^{T}\mathbf{R}_{\Delta {{\mathbf{b}}_{\theta }}}^{-1}\Delta {{\mathbf{b}}_{\theta }}+\Delta \mathbf{b}_{\tau }^{T}\mathbf{R}_{\Delta {{\mathbf{b}}_{\tau }}}^{-1}\Delta {{\mathbf{b}}_{\tau }},  \nonumber 
\end{align}
where ${{\mathbf{R}}_{\Delta {{\mathbf{A}}_{\theta }}}}$ and ${{\mathbf{R}}_{\Delta \mathbf{b}}}$ and represent the variances of $\text{vec}\left( \Delta {{\mathbf{A}}_{\theta }} \right)$ and $\Delta \mathbf{b}$ respectively.
By taking the partial derivatives with respect to each variable and setting the results to zero, we can obtain the following results.
\begin{align}
  \label{es2eq1}& \Delta {{\mathbf{b}}_{\tau }}={{{\mathbf{R}}_{\Delta {{\mathbf{b}}_{\tau }}}}{{\boldsymbol{\lambda }}_{\tau }}}/{2}\;, \\ \label{es2eq2}
 & \Delta {{\mathbf{b}}_{\theta }}={{{\mathbf{R}}_{\Delta {{\mathbf{b}}_{\theta }}}}{{\boldsymbol{\lambda }}_{\theta }}}/{2}\;, \\ \label{es2eq3}
 & \Delta {{\mathbf{a}}_{\theta }}=-{\mathbf{e}_{3}^{T}\mathbf{x}_2{{\mathbf{R}}_{\Delta {{\mathbf{A}}_{\theta }}}}{{\boldsymbol{\lambda }}_{\theta }}}/{2}\;, \\ \label{es2eq4}
 & {{\mathbf{A}}_{\tau }}\mathbf{x}_2-{{\mathbf{b}}_{\tau }}-\Delta {{\mathbf{b}}_{\tau }}=0, \\ \label{es2eq5}
 & {{\mathbf{A}}_{\theta }}\mathbf{x}_2+\Delta {{\mathbf{a}}_{\theta }}\mathbf{e}_{3}^{T}\mathbf{x}_2-{{\mathbf{b}}_{\theta }}-\Delta {{\mathbf{b}}_{\theta }}=0, \\ \label{es2eq6}
 & {{\mathbf{A}}_{\theta }}{{\boldsymbol{\lambda }}_{\theta }}+\boldsymbol{\lambda }_{\theta }^{T}\Delta {{\mathbf{a}}_{\theta }}{{\mathbf{e}}_{3}}+{{\mathbf{A}}_{\tau }}{{\boldsymbol{\lambda }}_{\tau }}=0. 
\end{align}
By substituting equation (\ref{es2eq1}) into (\ref{es2eq4}) and equations (\ref{es2eq2}) and (\ref{es2eq3}) into (\ref{es2eq5}), we can obtain
\begin{align}
  \label{es2eq7}& {{\boldsymbol{\lambda }}_{\tau }}=2\mathbf{R}_{\Delta {{\mathbf{b}}_{\tau }}}^{-1}\left( {{\mathbf{A}}_{\tau }}\mathbf{x}-{{\mathbf{b}}_{\tau }} \right), \\ 
 \label{es2eq8}& {{\boldsymbol{\lambda }}_{\theta }}=2{{\left( \mathbf{e}_{3}^{T}\mathbf{xe}_{3}^{T}\mathbf{x}_2{{\mathbf{R}}_{\Delta {{\mathbf{A}}_{\theta }}}}+{{\mathbf{R}}_{\Delta {{\mathbf{b}}_{\theta }}}} \right)}^{-1}}\left( {{\mathbf{A}}_{\theta }}\mathbf{x}_2-{{\mathbf{b}}_{\theta }} \right). 
\end{align}
For the convenience of subsequent processing, we define
$\tilde{x}\triangleq \mathbf{e}_{3}^{T}\mathbf{x}_2\mathbf{e}_{3}^{T}\mathbf{x}_2$. By 
substituting equation (\ref{es2eq7}) and (\ref{es2eq8}) into (\ref{es2eq6}), the estimate of $\mathbf{x}_2$ can be expressed as
\begin{align}\label{esx2}
\mathbf{\hat{x}_{2}}={{\left( {{\mathbf{B}}_{\theta }}{{\mathbf{A}}_{\theta }}+{{\mathbf{A}}_{\tau }}\mathbf{R}_{\Delta {{\mathbf{b}}_{\tau }}}^{-1}{{\mathbf{A}}_{\tau }} \right)}^{-1}}\left( {{\mathbf{B}}_{\theta }}{{\mathbf{b}}_{\theta }}+{{\mathbf{A}}_{\tau }}\mathbf{R}_{\Delta {{\mathbf{b}}_{\tau }}}^{-1}{{\mathbf{b}}_{\tau }} \right),
\end{align}
where ${{\mathbf{B}}_{\theta }}=\left( {{\mathbf{A}}_{\theta }}-\left[ \begin{matrix}
   \Delta {{\mathbf{a}}_{\theta }} & {{\mathbf{0}}_{K\times 2}}  \\
\end{matrix} \right] \right){{\left( \tilde{x}{{\mathbf{R}}_{\Delta {{\mathbf{A}}_{\theta }}}}+{{\mathbf{R}}_{\Delta {{\mathbf{b}}_{\theta }}}} \right)}^{-1}}$. We update the corresponding variables using an iterative method until the estimation results converge. The initial estimate can be obtained using least squares estimation, i.e., ${{\mathbf{\hat{x}}}_{\text{LS}}}={{\left( {{\mathbf{A}}^{T}_2}\mathbf{A}_2 \right)}^{\dagger }}{{\mathbf{A}}^{T}_2}\mathbf{b}_2$.
\begin{figure*}[!t]
	\normalsize
{\color{blue}\begin{align}\label{definationaxb}
  & {{\mathbf{A}}_{\tau }}\triangleq \left[ \begin{matrix}
   -2{{x}_{\text{I},1}} & -2{{y}_{\text{I},1}} & 1  \\
   \cdots  & \cdots  & \cdots   \\
   -2{{x}_{\text{I},K}} & -2{{y}_{\text{I},K}} & 1  \\
\end{matrix} \right],{{\mathbf{A}}_{\theta }}\triangleq \left[ \begin{matrix}
   \tan {{{\hat{\theta }}}_{1}} & -1 & 0  \\
   \cdots  & \cdots  & \cdots   \\
   \tan {{{\hat{\theta }}}_{K}} & -1 & 0  \\
\end{matrix} \right],\Delta {{\mathbf{A}}_{\theta }}\triangleq \left[ \begin{matrix}
   \Delta {{\mathbf{a}}_{\theta }} & {{\mathbf{0}}_{K\times 2}}  \\
\end{matrix} \right]=\left[ \begin{matrix}
   \tan {{\varepsilon }_{1}} & 0 & 0  \\
   \cdots  & \cdots  & \cdots   \\
   \tan {{\varepsilon }_{K}} & 0 & 0  \\
\end{matrix} \right],\nonumber \\ 
 & {{\mathbf{b}}_{\tau }}\triangleq \left[ \hat{\tau }_{1}^{2}{{c}^{2}}-\left( x_{\text{I},\text{1}}^{2}+y_{\text{I},\text{1}}^{2} \right),\cdots ,\hat{\tau }_{K}^{2}{{c}^{2}}-\left( x_{\text{I},K}^{2}+y_{\text{I},K}^{2}\right) \right],\Delta {{\mathbf{b}}_{\tau }}\triangleq \left[ 2{{\tau }_{1}}{{\varepsilon }_{1}}c^2,\cdots ,2{{\tau }_{K}}{{\varepsilon }_{K}}c^2 \right],\\ 
 & {{\mathbf{b}}_{\theta }}\triangleq \left[ {{x}_{\text{I},1}}\tan{{{\hat{\theta }}}_{1}}-{{y}_{\text{I},1}},\cdots ,{{x}_{\text{I},K}}\tan{{{\hat{\theta }}}_{k}}-{{y}_{\text{I},K}} \right],\Delta {{\mathbf{b}}_{\theta }}\triangleq \left[ {{x}_{\text{I},1}}\tan{{\varepsilon }_{1}},\cdots ,{{x}_{\text{I},K}}\tan{{\varepsilon }_{K}} \right], \nonumber  
\end{align}}
	\hrulefill
	\vspace*{1pt}
\end{figure*}

\subsection{Stage 3} 

After completing the second stage of estimation, the error between the estimated values and the true values can be modeled as
\begin{align}
  & {{x}_{\text{T}}}={{{\hat{x}}}_{\text{T,2}}}+{{\varepsilon }_{x}},\label{1}\\ 
 & {{y}_{\text{T}}}={{{\hat{y}}}_{\text{T,2}}}+{{\varepsilon }_{y}},\\ 
 & x_{\text{T}}^{2}+y_{\text{T}}^{2}=\hat{s}+{{\varepsilon }_{s}},  
\end{align}
where ${{{\hat{x}}}_{\text{T,2}}}$, ${{{\hat{y}}}_{\text{T,2}}}$, and ${{{\hat{s}}}_{\text{T}}}$ represent the estimations of ${{x}_{\text{T}}}$, ${{y}_{\text{T}}}$, and $x_{\text{T}}^{2}+y_{\text{T}}^{2}$ based on the second stage of estimation and  ${{\varepsilon }_{x}}$, ${{\varepsilon }_{y}}$, and ${{\varepsilon }_{s}}$ represent the corresponding estimation errors. 

In the third stage, we utilize the location information contained in $s$ to refine the solution obtained in the first stage. This leads to the following system of equations
\begin{align}
\underbrace{\left[ \begin{matrix}
   1 & 0  \\
   0 & 1  \\
   1 & 1  \\
\end{matrix} \right]}_{{{\mathbf{A}}_{3}}}\underbrace{\left[ \begin{aligned}
  & x_{\text{T}}^{2} \\ 
 & y_{\text{T}}^{2} \\ 
\end{aligned} \right]}_{{{\mathbf{x}}_{3}}}=\underbrace{\left[ \begin{aligned}
  & \hat{x}_{\text{T,2}}^{2} \\ 
 & \hat{y}_{\text{T,2}}^{2} \\ 
 & {\hat{s}} \\ 
\end{aligned} \right]}_{{{\mathbf{b}}_{3}}}+\underbrace{\left[ \begin{aligned}
  & 2{{x}_{\text{T}}}{{\varepsilon }_{x}} \\ 
 & 2{{y}_{\text{T}}}{{\varepsilon }_{y}} \\ 
 & {{\varepsilon }_{s}} \\ 
\end{aligned} \right]}_{\Delta {{\mathbf{b}}_{3}}}.
\end{align}
and the least squares estimation for ${{{\mathbf{x}}_{3}}}$ can be expressed as
\begin{align}\label{esx3}
{{\mathbf{\hat{x}}}_{3}}={{\left( \mathbf{A}_{3}^{T}\mathbf{Q}_{3}^{-1}{{\mathbf{A}}_{3}} \right)}^{-1}}\mathbf{A}_{3}^{T}\mathbf{Q}_{3}^{-1}{{\mathbf{b}}_{3}}
\end{align}
where ${{\mathbf{Q}}_{3}}=\text{diag}\left( 4\hat{x}_{\text{T}}^{2},4\hat{y}_{\text{T}}^{2},\hat{s} \right)\text{conv}\left( {{{\mathbf{\hat{x}}}}_{2}} \right)\text{diag}\left( 4\hat{x}_{\text{T}}^{2},4\hat{y}_{\text{T}}^{2},\hat{s} \right)$. After obtaining the estimation ${{\mathbf{\hat{x}}}_{3}}$ from the second stage, we take the square root of $\hat{x_{\text{T}}^{2}}$ and $\hat{y_{\text{T}}^{2}}$ in ${{\mathbf{\hat{x}}}_{3}}$ , ensuring that the signs are consistent with those in the first stage, and then complete the final location estimation. and the entire three-stage localization algorithm is summarized in \textbf{Algorithm 3}.

\begin{algorithm}[t]
	\caption{Proposed Three-Stage  Localization Algorithm.} 
	\begin{algorithmic}[1]
		\State$\textbf{Input}$: Initialize $\hat{\boldsymbol{\tau }}$
and $\hat{\boldsymbol{\theta }}$
threshold ${{\varepsilon }_{2}}$, ${{\varepsilon }_{3}}$ and iteration index $r=1$
\State$\textbf{Stage 1}$: 
 \State Update ${{\mathbf{\hat{x}}_{1}}}$ according to (\ref{esx1}) .
\State$\textbf{Stage 2}$: 
		\Repeat:
        \State Update $\tilde{x}$ according to its definition. 
        \State Update ${{\mathbf{\hat{x}}_{2}}}$ according to (\ref{esx2}) . 
\Until 
$\left\| {{{\mathbf{\hat{x}}}}^{r+1}_{2}}-{{{\mathbf{\hat{x}}}}^{r}_{2}} \right\|_{2}^{2}\le {{\varepsilon }_{3}}$.
\State$\textbf{Stage 3}$: 
\State ${{\hat{x}}_{\text{T,2}}}={{\left[ {{{\mathbf{\hat{x}}}}_{\text{2}}} \right]}_{1}},{{\hat{y}}_{\text{T,2}}}={{\left[ {{{\mathbf{\hat{x}}}}_{2}} \right]}_{2}},\hat{s}={{\left[ {{{\mathbf{\hat{x}}}}_{2}} \right]}_{3}}$.
\State Calculate ${{\mathbf{\hat{x}}}_{3}}$ according to (\ref{esx3}).
\State ${{\hat{x}}_{\text{T,3}}}=\pm \sqrt{{{\left[ {{{\mathbf{\hat{x}}}}_{3}} \right]}_{1}}},{{\hat{y}}_{\text{T,3}}}=\pm \sqrt{{{\left[ {{{\mathbf{\hat{x}}}}_{3}} \right]}_{2}}}$ and the
signs are consistent with ${{\hat{x}}_{\text{T,2}}},{{\hat{y}}_{\text{T,2}}}$.
\State$\textbf{Output}$: $\mathbf{\hat{x}}=\left[ {{{\hat{x}}}_{\text{T,3}}},{{{\hat{y}}}_{\text{T,3}}} \right]$.
	\end{algorithmic}
\end{algorithm}

\section{Simulation Results}
In this section, we validate the superiority of the proposed multi-IRS collaborative localization model and the accuracy of the angle and location estimation algorithms through numerical simulations. We consider a two-dimensional coordinate system in this section. The BS is located at (0, 0) m while $K=3$ IRSs are located at (10, 50) m, (10, -50) m, and (50, 0) m. {\color{blue} A single target is randomly distributed within the rectangular area defined by $0<x<10$ m and $0<y<10$ m.
A key aspect of our simulation setup is the configuration of the IRS elements. In our simulations, we adopt a 
random phase configuration, where the phase shift of each element is chosen independently and uniformly from $\left[ 0,2\pi  \right)$. In the setup, each IRS is a Uniform Linear Array (ULA) placed vertically with its broadside oriented towards the origin. The target's Radar Cross Section ($\kappa $), listed in Table I, is modeled as a constant value independent of the viewing angle.}
Other main system parameters are listed in Table I. 
\begin{table}[t]
\centering
\small
\caption{Simulation Parameters}
\scalebox{0.88}{
\begin{tabular}{|c|c|c|c|c|c|}
\hline Parameter & Value & Parameter & Value & Parameter & Value\\
\hline$N_t$ & $50$ & ${N}$ & $10$ & $M$ & $10$\\
\hline$W$ & $50MHz$ & ${L}$ & $100$ & $\kappa $ & $7 \text{ dBsm}$\\
\hline${{P}_{\max}}$ & $50\text{dBmW}$ & $\sigma_{l}^{2}$ & $-100\text{dBm}$ & $\lambda$ & $0.3\text{m}$\\
\hline
\end{tabular}}
\end{table}
\subsection{CRB Performance for Localization System}
The accuracy of the localization system within the region is characterized by calculating the average CRB of localization estimation of multiple random target, i.e.,   {\color{blue}       
\begin{align}
{{\text{C}}_{\text{ave}}}=\frac{1}{{{N}_{\text{trials}}}}\sum\limits_{i=1}^{{{N}_{\text{trials}}}}{\text{C}\left( {{\mathbf{l}}_{\text{T},i}} \right)},
\end{align}
where we set ${{N}_{\text{trials}}}$ is the total number of Monte Carlo trials, set to 100; $\text{C}\left( {{\mathbf{x}}_{\text{T},i}} \right)$ is the CRB  for a single target at a random location ${{\mathbf{l}}_{\text{T},i}}$.} We compare the proposed system design with the following schemes:
\begin{itemize}
\item \textbf{Angle-based:} The system only estimates the angle and performs localization using the angle information.

\item \textbf{Delay-based:} The system only estimates the delay and performs localization using the delay information.

\item \textbf{No collaboration:} All semi-passive IRSs employ frequency-division or code-division techniques for the independent reflection of incident signals and reception of echo signals, thereby facilitating independent localization.

\item \textbf{Single IRS:} All the reflective elements and sensors are deployed at the location of the first IRS.

\end{itemize}
\subsubsection{CRB Versus Number of Reflective Elements}
\begin{figure}
\centerline{\includegraphics[width=8cm]{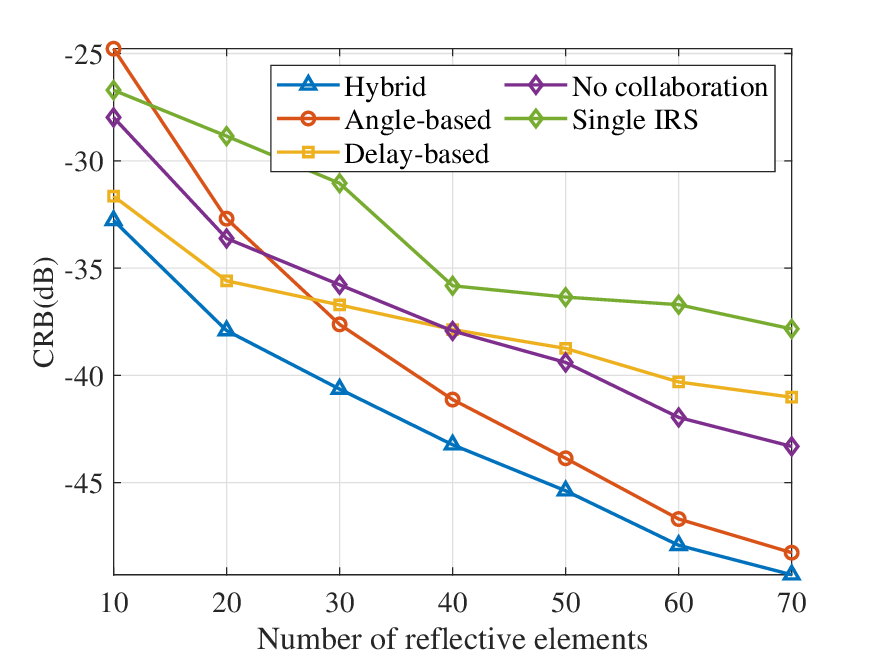}}
	\caption{CRB versus the number of reflective elements.}\label{s_n}
\end{figure}

In Fig. \ref{s_n}, we compare the average CRB for location estimation versus the number of reflective elements. First, the performance of all schemes improves as the number of reflective elements increases, which can be attributed to two main reasons. One reason is that the transmitter units provide beamforming gain of $\mathcal{O}\left( {N} \right)$ (in this model, we do not consider the optimal IRS coefficient design, which could provide an $\mathcal{O}\left( {N^2} \right)$ gain. Our model assumes random IRS coefficients, resulting in an $\mathcal{O}\left( {N} \right)$ gain), which positively affects the estimation of delay, AOA, and AOD. The CRB scales down with $N$ approximately in the the order of $1/N$. Another reason is that for AOD estimation, the increase in reflective elements leads to a higher dimensionality of the corresponding steering vector, thereby enhancing spatial resolution.


\subsubsection{CRB Versus Number of Sensors}
\begin{figure}
\centerline{\includegraphics[width=8cm]{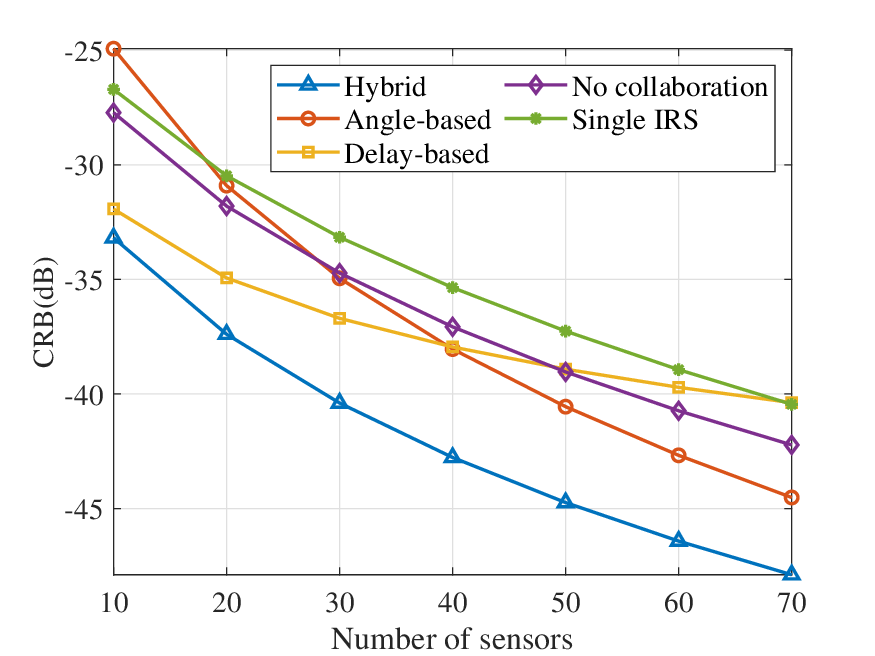}}
	\caption{CRB versus the number of sensors.}\label{s_m}

\end{figure}

In Fig. \ref{s_m}, we compare the CRB for location estimation of the single target versus the number of transmitting antennas, Based on Fig. \ref{s_n} and Fig. \ref{s_m}, we observe that multiple semi-passive IRS collaborative localization can improve accuracy by approximately 7 dB, and this value remains unchanged regardless of the number of reflective elements or sensors. We also find that when the number of reflective elements or sensors is low, the delay information plays a dominant role in localization, while at higher numbers of reflective elements or sensors, the angle information becomes dominant. This is because, in the case of random IRS coefficients, the Fisher information provided by delay information is approximately $\mathcal{O}\left( {MN}\right)$, while the Fisher information provided by AOA and AOD is approximately $\mathcal{O}\left( {M^3N} \right)$ and $\mathcal{O}\left( {MN^3} \right)$, respectively. Therefore, for the design of IRS-assisted localization systems, the choice of direct estimators can be determined based on the number of reflective elements, the number of sensors, and the bandwidth $W$ of the sensing signal. This choice will influence the complexity of signal processing. Finally, we observe a significant performance difference between traditional single-IRS and multi-IRS systems, which remains unchanged regardless of the number of reflective elements or sensors. To investigate the main reasons behind this difference, we will explore the impact of distance in the next subsection.

\subsubsection{CRB Versus Coverage area}
\begin{figure}
\centerline{\includegraphics[width=8cm]{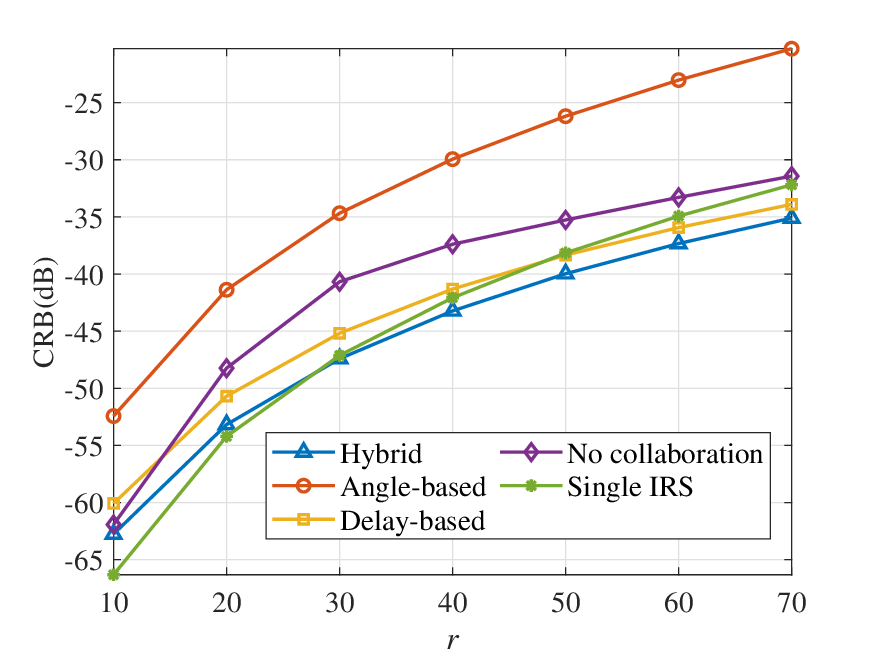}}
	\caption{CRB versus the coverage area.}
	\label{s_d}
\end{figure}

In Fig. \ref{s_d}, we change the target area for localization to a circle with the first semi-passive IRS as the center and a radius of $r$ and compare the average CRB for location estimation versus $r$. 
it is observed that When $r$ is less than 30m, the performance of the single-IRS system is the best. However, as $r$ increases, the performance advantage shifts to the multi-IRS system. This is because at larger distances, the benefit of mitigating path loss through distributed deployment outweighs the drawback of splitting system resources. {\color{blue}This observation reveals the distinct application scenarios for each deployment strategy: a single-IRS deployment is more suitable for localization tasks in "hotspot" areas where the target is relatively static and at a close range, whereas a multi-IRS collaborative deployment is better adapted for localization over larger areas, significantly enhancing the system's coverage and robustness.}
Furthermore, it is interesting to
note that the collaborative gain varies with $r$ and does not remain the same as in the previous two simulations. This is because, as $r$ changes, the degree to which multiple IRSs are involved in the system also changes. Specifically, when the distance between a certain IRS and the target is much smaller than the distances of the other IRSs to the target, that IRS dominates the localization, resulting in a lower collaborative gain. On the other hand, when the distances from all IRSs to the target are roughly the same, the collaborative gain is maximized. Furthermore, the collaborative gain increases as the number of IRSs increases, because cooperative signal processing can expand the number of observation samples from $K$ to $K^2$.

\subsection{Performance of Angle Estimation Algorithm}
In this section, we select the reflection signal from the IRS with ${{\mathbf{{S}}}_{k}}=N$ for simulation. The algorithm proposed in Sections IV is compared with several classical angle estimation algorithms, and the MSE of $K$ angles is calculated to verify the accuracy of the angle estimation.
\subsubsection{MSE Versus transmitting power at the BS}
\begin{figure}
\centerline{\includegraphics[width=8cm]{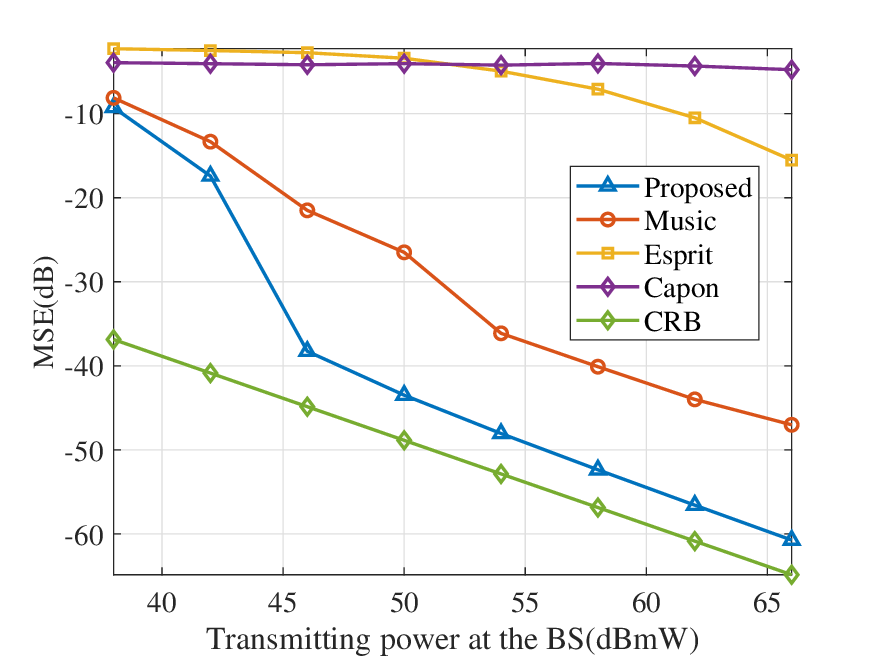}}
	\caption{MSE versus the transmitting power at the BS.}
	\label{es_ap}
\end{figure}

Fig. \ref{es_ap} shows the average MSE versus transmitting power at the BS. It can be observed that the estimation error decreases linearly with increasing power. The proposed estimation algorithm achieves the best performance compared to other benchmarks, with only a 4 dB gap from the CRB. The Capon algorithm is sensitive to noise and thus fails to accurately estimate angles under the SNR conditions set in the simulation. The ESPRIT algorithm can perform angle estimation only at relatively high SNR levels. Although the MUSIC algorithm is capable of estimation under lower SNR, it exhibits a gap of more than 10 dB from the CRB. Since this paper does not consider the design of optimal IRS coefficients, the required base station transmission power is relatively high. In practice, with proper beamforming design, the required transmission power can be significantly reduced.

\subsection{Performance of Localization Estimation Algorithm}
We generate error-prone estimates of $\hat{\boldsymbol{\tau }}$
and $\hat{\boldsymbol{\theta }}$ with variances corresponding to the derived CRB of the cascaded delay and angle.  Then, the estimation of  based on $\hat{\boldsymbol{\tau }}$
and $\hat{\boldsymbol{\theta }}$ is completed using the three-stage algorithm proposed in Section V, and compared with the following schemes.        
\begin{itemize}
\item \textbf{Two-stage:} Only the first two stages of the three-stage algorithm are used, meaning the information in the second-order terms is ignored.

\item \textbf{WLS:} After estimating the segmented delay in the first stage, WLS estimation is applied to estimate the location.

\item \textbf{LS:} After estimating the segmented delay in the first stage, LS estimation is applied to estimate the location.

\item \textbf{CRB:} The CRB derived in Section III.

\end{itemize}

\subsubsection{MSE Versus transmitting power at the BS}
\begin{figure}
\centerline{\includegraphics[width=8cm]{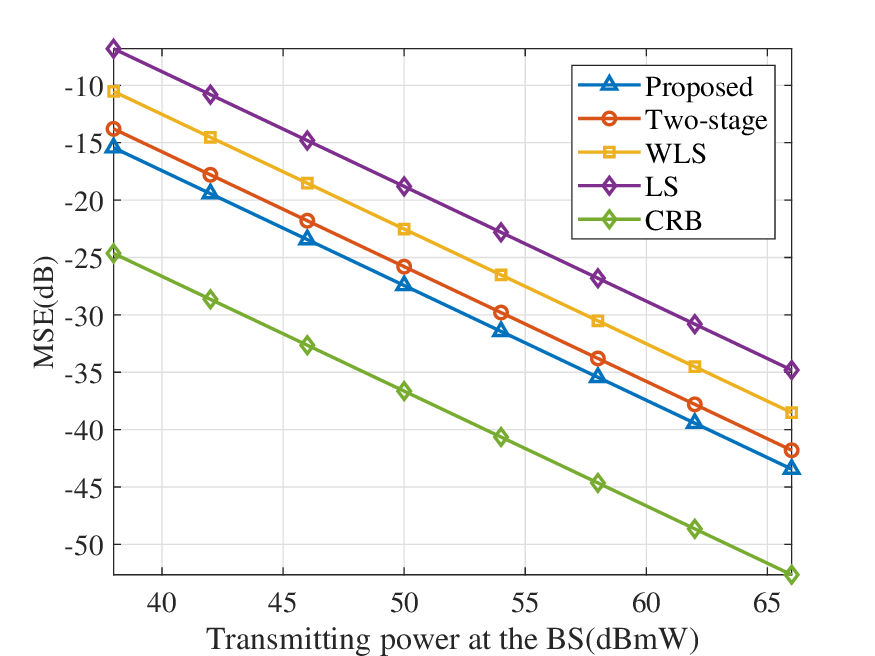}}
	\caption{MSE versus the transmitting power at the BS.}
	\label{es_p}
\end{figure}

Fig. \ref{es_p} shows the average MSE versus transmitting power at the BS. The MSE differences among different estimation schemes are significant. Specifically, the LS estimation has the highest MSE because it completely ignores the variance information of the errors in the coefficient matrix $\mathbf{A_2}$ and the observation sample $\mathbf{b_2}$, as well as the quadratic information related to the location contained in the observation samples. Compared to LS, WLS takes into account the variance information of the errors in the observation sample $\mathbf{b_2}$, leading to an approximately 4 dB reduction in the estimated MSE. Building upon this, our proposed algorithm considers the errors in the coefficient matrix $\mathbf{A_2}$ in the second stage and the quadratic information of the location included in $\mathbf{b_2}$ in the third stage, further reducing the estimation MSE by 4 dB and 2 dB, respectively. Finally, we found that the performance improvement gained by addressing the three major challenges not considered in LS estimation remains unaffected by changes in the signal-SNR. Therefore, in practical locationing algorithms, these challenges should be addressed in both high and low SNR scenarios.

\subsubsection{MSE Versus the distance}
\begin{figure}
\centerline{\includegraphics[width=8cm]{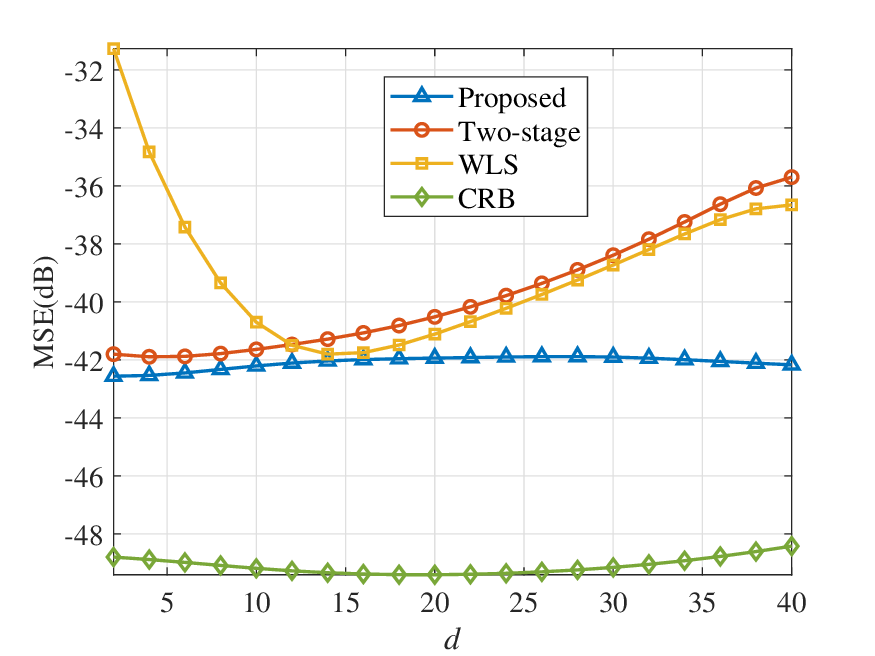}}
	\caption{CRB versus the distance.}
	\label{es_d}
\end{figure}

In this subsection, we explore the impact of IRS deployment on estimation performance by changing the coordinates of the first two IRS location from $\left( 10,-50 \right)$ and $\left( 10,50 \right)$ to $\left(d,-50 \right)$ and $\left(d,50 \right)$, respectively. Fig. \ref{es_d} shows the average MSE  versus $d$. It can be observed that only the estimation error of the proposed scheme experiences minimal fluctuations, similar to the CRB, as the IRS deployment changes within the given range.  When $d$ is small, the error of the WLS algorithm increases because the path loss of the system signal increases as $d$ decreases, particularly when $d<10$. Specifically, the horizontal distance between the first two IRSs and the base station is $d$, and the horizontal distance between the first two IRSs and the target is $x-d$. The echoed signal travels through the latter distance twice. The necessity of the third-stage algorithm increases as $d$ increases. This is because the variance of the quadratic term estimation related to the location decreases as the absolute value of the IRS coordinates increases. When the variance of the quadratic term estimation becomes smaller, the correction of the coarse estimate by the third-stage algorithm becomes more accurate.

\section{Conclusions}
This paper presents a novel collaborative hybrid localization system using multi-IRS to estimate target location through joint time delay and angle estimation. Echo signals from IRS reflective elements are processed by sensors to estimate time delay and angle parameters. The FIM and CRB are derived for delay, AOA, and AOD estimation in. Efficient algorithms for angle and location estimation are designed, including a convex optimization approach using ADMM for AOA/AOD estimation and a three-stage algorithm combining least squares methods for location estimation. Simulations show that the system provides significant benefits, especially in low signal-to-noise ratio conditions.

\bibliographystyle{IEEEtran}
\bibliography{reference}
\end{document}